\documentclass[amsmath,amssymb,aps,prb,floatfix,twocolumn,10pt]{revtex4-2}

\usepackage{graphicx}
\usepackage{dcolumn}
\usepackage{bm}
\usepackage{soul}
\usepackage{subfigure}
\usepackage{amsthm}
\usepackage{amsfonts}
\usepackage{braket}
\usepackage{comment}
\usepackage{enumitem}
\usepackage{float}
\usepackage{dsfont}
\usepackage{xcolor}
\usepackage{soul}
\usepackage{hyperref}
\usepackage[capitalize]{cleveref} 
\hypersetup{
    breaklinks=true,
    colorlinks=true,
    linkcolor=blue,
    urlcolor=blue,
    citecolor=blue
}
\def\nn{\nonumber\\}
\def\be{\begin{equation}}
\def\ee{\end{equation}}
\def\bla{{\boldsymbol{\lambda}}}
\def\blambda{{\boldsymbol{\lambda}}}
\def\bmu{{\boldsymbol{\mu}}}
\newcommand{\fab}[1]{{\color{red}\ifmmode\text{\footnotesize(FE) #1}\else\footnotesize{(FE) #1}\fi}}

\begin{document}

\preprint{APS/123-QED}

\title{Finite temperature single-particle Green's function in the Lieb-Liniger model}
\author{Riccardo Senese}
\affiliation{Rudolf Peierls Centre for Theoretical Physics, University of Oxford, Oxford OX1 3PU, United Kingdom}
\author{Fabian H.L. Essler}%
\affiliation{Rudolf Peierls Centre for Theoretical Physics, University of Oxford, Oxford OX1 3PU, United Kingdom}

\date{\today}

\begin{abstract}
We develop a Monte Carlo sampling algorithm to numerically evaluate the Lehmann representation for the finite temperature single-particle Green's function in the repulsive Lieb-Liniger model. This allows us to determine the spectral function in the full range of temperatures and interactions, as well as in generalized Gibbs ensembles. We test our results against known results for dynamics at infinite interaction strength and static correlators, and find excellent agreement. 
\end{abstract}

\maketitle



\section{Introduction}
The Lieb-Liniger (LL) model is a key paradigm for interacting many-particle quantum systems in low dimensions. It has played a central role in the development of the theory of quantum integrability \cite{korepin1993quantum_inverse,lieb1963exact_1,lieb1963exact_2,yang1969thermodynamics,slavnov1989calculation,korepin1990time_dependent,slavnov1990nonequal,caux2006dynamical,caux2007one_particle,kitanine2012form,imambekov2012one_dimensional,kormos2013interaction,de_nardis2014solution,kozlowski2015large,piroli2016multiparticle,doyon2017drude,de_nardis2019diffusion,granet2020systematic,granet2021low,granet2021systematic}
and over the last decade and a half has been studied in great detail in cold atom experiments, both in \cite{kinoshita2004observation,paredes2004tonks,fabbri2015dynamical,meinert2015probing,fang2016momentum} and out of \cite{kinoshita2006quantum,hofferberth2007non,cheneau2012light,bouchoule2022generalized} equilibrium. In spite of the huge progress made, there remain important questions that have stubbornly resisted theoretical progress. In particular, the evaluation by either analytic or numerical means of dynamical correlation functions for large numbers of particles at general temperatures and interaction strengths has remained an open problem. Integrability provides an efficient representation of energy eigenstates, which can be leveraged to obtain Lehmann representations for dynamical correlators that are much simpler than in generic many-particle systems. However, a remaining fundamental obstacle is that these involve sums over “intermediate” eigenstates whose number scales \emph{exponentially} with system size. This problem is particularly acute for the two-point dynamical correlation function of the Bose field, which is the focus of our work. It is possible to express two-point correlators in terms of operator-valued Fredholm determinants \cite{korepin_correlation_1991, kojima1997determinant, kojima_completely_1997, its_riemann-hilbert_1999, slavnov_integral_1999}, but these have so far resisted further analysis by either numerical or analytical means. These difficulties motivated the development of sophisticated numerical approaches to \emph{explicitly} perform the Lehmann sums. However, these algorithms are restricted to zero \cite{caux2007one_particle, caux2009correlation, cheng_exact_2025} or extremely low \cite{cheng2025one_body} temperatures and small-to-intermediate system sizes. We show
that in order to treat arbitrary temperatures, generalized Gibbs ensembles, the full range of repulsive interactions, and fairly large particle numbers, a stochastic sampling of the huge number of \emph{relevant} eigenstates contributing to the Lehmann sum is required. We propose an algorithm that enables us to obtain accurate results in regimes that have been, until now, beyond the reach of classical computational methods.

\section{Model and Green's function}

We consider the LL model \cite{lieb1963exact_1, lieb1963exact_2} of bosons with repulsive delta-function interactions on a ring of length $L$ (we set $\hbar = 2m = 1$)
\begin{equation}
\label{eq:HsecondQ}
    H = \int_0^L\!\! dx \Big[ \phi^\dagger(x) [-\partial_{x}^2 - h] \phi(x) + c  \big(\phi^\dag(x)\big)^2\big(\phi(x)\big)^2 \Big] .
\end{equation}
Here $c>0$ is the interaction strength, $h$ a chemical potential and $\phi(x)$ is a complex bosonic field obeying commutation relations $[\phi(x),\phi^\dag (y)]=\delta(x-y)$. The Hamiltonian \eqref{eq:HsecondQ} has two non-interacting limits: free bosons for $c=0$ and free fermions in the limit $c\rightarrow\infty$ \cite{girdardeau1960relationship, creamer1980quantum}. 

By virtue of its integrability the LL model has an extensive number
of mutually compatible conservation laws $Q^{(n)}$
\cite{korepin1993quantum_inverse,davies_higher_1990}. The Hamiltonian \eqref{eq:HsecondQ} is $H=Q^{(2)}-hQ^{(0)}$, where $Q^{(0)}$ is the particle number operator.
Simultaneous eigenstates $\ket \bla$ of $\{Q^{(n)}\}$ in the $N$-particle sector are parametrized by a set of $N$ real “rapidities” $\bla=(\lambda_1, \ldots, \lambda_N)$.
The eigenvalues of $Q^{(n)}$ on the states $|\bla\rangle$ are $q^{(n)}_{\boldsymbol{\lambda}}=\sum_{i=1}^N\lambda_i^n$, and in particular the energy and momentum eigenvalues are given by
\be
E_{\bla}=\sum_{j=1}^N(\lambda_j^2-h)\ ,\quad P_{\bla}=\sum_{j=1}^N\lambda_j\ .
\ee
On a ring of length $L$ the rapidities fulfil the quantization conditions (“Bethe equations”)
\cite{lieb1963exact_1}
\begin{equation}
\label{eq:logBetheEq}
    L \lambda_j + \sum_{k=1}^N \theta(\lambda_j - \lambda_k) = 2 \pi I_j \qquad j = 1, \ldots, N \ .
\end{equation}
Here $\theta(x)=2 \arctan(x/c)$ and $\{I_j\}$ is a set of distinct integers (half-odd integers) for $N$ odd (even). The sets $\{ I_j\}$ are in one-to-one correspondence with the eigenstates $\ket \bla$. The objective of this work is to evaluate the dynamical two-point function of the Bose field in an energy eigenstate $|\bla\rangle$
\begin{equation}
\label{eq:Cxt}
    C(x,t) = \braket{\bla|\phi^\dag(x,t) \, \phi(0,0)|\bla} \ .
\end{equation}
The limit $N,L\to\infty$, $n=N/L=\text{const}$, is to be taken at fixed $x,t$. We consider two cases: (i) a microcanonical average where $|\bla\rangle$ is a thermal state at non-zero temperature $T$ \cite{korepin1993quantum_inverse}; (ii) a generalized microcanonical average \cite{cassidy2011generalized,caux2013time} corresponding to a generalized Gibbs ensemble (GGE) \cite{rigol2007relaxation,ilievski2015complete} with density matrix $\hat{\rho}=Z^{-1}\exp(-\sum_n\beta_nQ^{(n)})$.
Details of the construction of $\ket{\bla}$ are given in Appendix \ref{appendix:GME}. 

Inserting $\mathds{1}=\sum_\bmu\ket{\bmu}\bra{\bmu}$ in \eqref{eq:Cxt} gives
\begin{align}
\label{eq:Cxtdef}
C(x,t) &=  \sum_\bmu g_{\bla,\bmu}(x,t)\,  |\braket{\bmu|\phi(0,0)|\bla}|^2 \ ,\nn
g_{\bla,\bmu}(x,t) &= e^{i(E_{\bla}-E_{\bmu})t - i(P_\bla-P_\bmu)x} \ .
\end{align}
Exact expressions for $\braket{\bmu|\phi(0,0)|\blambda}$ are known \cite{kojima1997determinant, caux2007one_particle, piroli2015exact_formulas} (see Appendix \ref{appendix:formfactors}) and they prove that the form factors $\braket{\bmu|\phi(0,0)|\blambda}$ are non-vanishing for any eigenstate $\ket \bmu$ of $N-1$ particles. In contrast to systems satisfying the Eigenstate Thermalization Hypothesis, where matrix elements like $\braket{\bmu|\phi(0,0)|\blambda}$ are all exponentially small in $L$ \cite{srednicki1999approach, d_alessio2016from}, their structure is considerably more complex in integrable systems \cite{essler2024statistics}. In particular, only a sub-entropic yet exponentially large in system size fraction of eigenstates $\ket \bmu$ contributes to the Lehmann sum \eqref{eq:Cxtdef} for $L\to\infty$. The necessity of accounting for an exponentially large number (in $L$) of terms in (\ref{eq:Cxtdef}) requires a different approach compared to the one applicable at zero, or very low \footnote{We remark that in defining what we mean by a low temperature it is important to take into account also the system size $L$. For large $L$ (and density $n=N/L=1$), we have in mind $T \ll 1$. We notice, however, that for $L\approx 60$-$80$, Ref.~\cite{cheng2025one_body} managed to obtain results for $T=1.5$. In this parameter regime the reference state $\ket{\boldsymbol{\lambda}}$ differs from the ground-state ($T=0$) of the model by \emph{very few} particle-hole excitations in the bulk of the Fermi sea. This is why we consider also this case, effectively, as a very low temperature.}, temperature \cite{caux2007one_particle,caux2009correlation, de_klerk2023improved,cheng2025one_body, cheng_exact_2025}. In this work we develop such an approach.

\section{Monte Carlo sampling scheme}
In order to evaluate the Lehmann representation (\ref{eq:Cxtdef})
we employ Markov chain Monte Carlo (MCMC) sampling \cite{metropolis1953equation, hastings1970monte, rubinstein2016simulation,gilks1995markov,feller1957introduction} of the relevant form factors. We note that the idea of estimating sums over eigenstates by MCMC previously appeared in Refs \cite{gritsev2010exact, buccheri2011structure,faribault2013integrability,faribault2013spin,burovski_momentum_2014,alba2015simulating,alba_quench_2016,gamayun_impact_2018,bouchoule2020effect, zhang_monte_2024} \footnote{We note that other Monte Carlo approaches have been used in the context of integrability or coupled with the Bethe ansatz technology, see e.g.~\cite{gu2005numerical, de_rosi_hole-induced_2022, de_rosi_correlation_2023}, but these do not involve sampling over form factors or overlaps and so are significantly different.}, in several contexts. However, the evaluation of dynamical finite-temperature correlation functions and the systematic analysis of the accuracy and applicability of such Monte Carlo approaches have not been pursued to date. We rewrite \eqref{eq:Cxtdef} as
\begin{equation}
\label{eq:Cxt2}
\begin{aligned}
C(x,t) &= Z_{\blambda} \sum_\mu g_{\blambda,\bmu}(x,t) \frac{e^{-M_{\blambda,\bmu}}}{Z_\blambda} \ ,\\
M_{\blambda,\bmu}&=-\ln(|\braket{\bmu|\phi(0,0)|\bla}|^2) \ ,
\end{aligned}
\end{equation}
where the “partition function” $Z_\blambda$ simply corresponds to
\begin{equation}
\begin{aligned}
\label{eq:partitionfunctionZlambda}
Z_\blambda &= \sum_\bmu e^{-M_{\blambda,\bmu}}
= \braket{\blambda | \phi^\dag(0,0) \, \phi(0,0)|\blambda} = n \ .
\end{aligned}
\end{equation}
To draw an analogy we can think of $\mathcal{P}_\blambda(\bmu)\equiv\exp(-M_{\blambda,\bmu})/Z_{\blambda}$ as a normalized Gibbs distribution, where $M_{\blambda,\bmu}$ plays the role of an “energy”. The key realization is that eigenstates $\ket{\bmu_\ell}$ can be sampled according to $\mathcal{P}_{\bla}(\bmu)$ via MCMC, yielding the following estimate
\begin{equation}
\label{eq:Cxt_estimate}
    C(x,t) \ \rightarrow  \ \frac{Z_\blambda}{\ell_{\rm max}}\sum_{\ell=1}^{\ell_{\rm max}}g_{\blambda,\bmu_\ell}(x,t) \ .
\end{equation}
The sampling is achieved by means of the Metropolis-Hastings algorithm \cite{metropolis1953equation, hastings1970monte} with single-integer updates, \emph{cf.}~\cref{eq:logBetheEq}, as detailed in Appendix \ref{appendix:detailsofMCMC}.
\begin{figure}[b]
    \centering
    \includegraphics[width=0.95\linewidth]{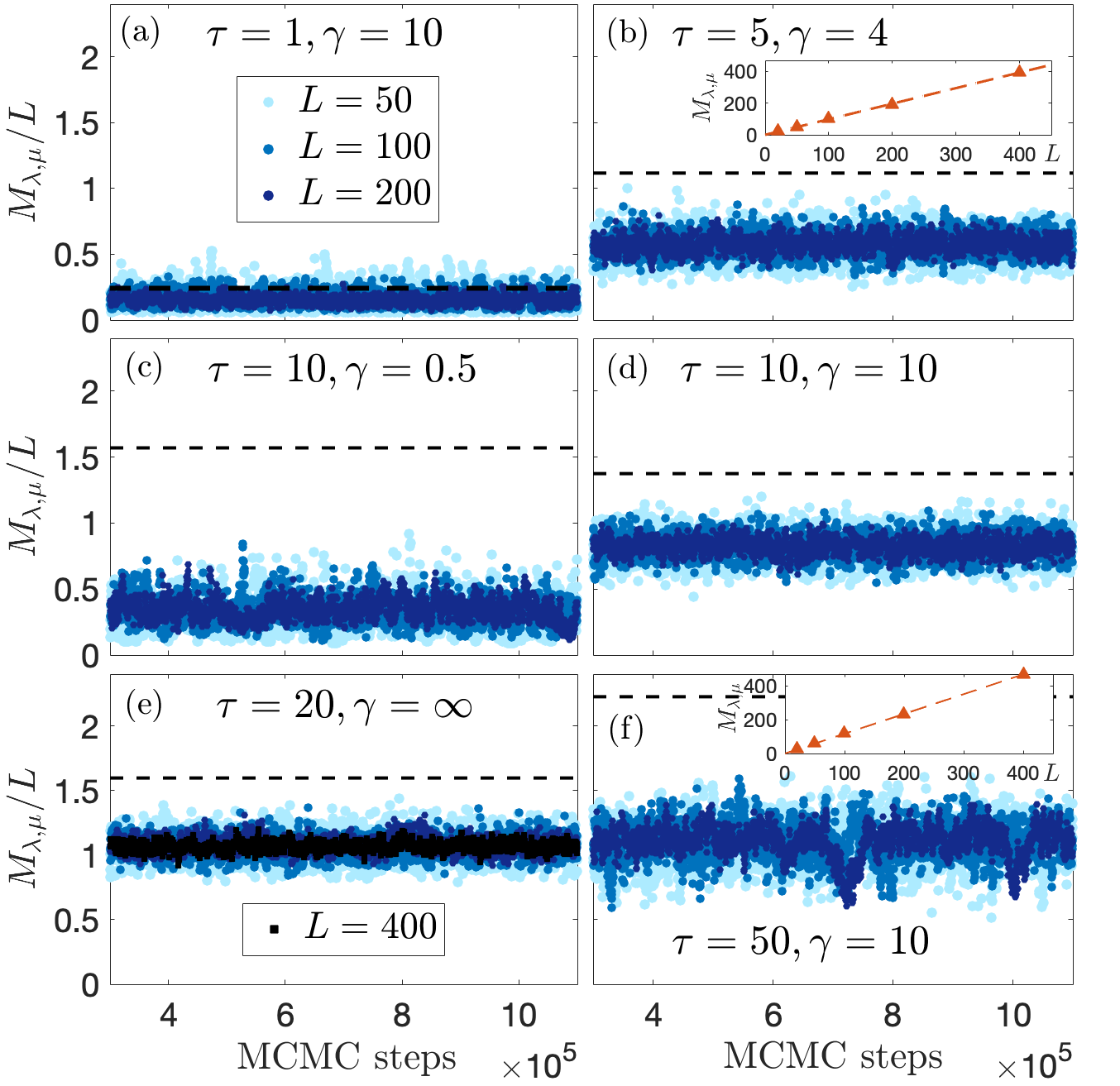}
    \caption{$M_{\blambda,\bmu}/L$ sampled in a single MCMC run, at different couplings $\gamma$, temperatures $\tau$ and sizes $L$. We plot one value every 3000 MCMC steps. Dashed lines: entropy densities $s[\rho]$ associated with the thermal macrostates (Appendix \ref{appendix:GME}). The insets in (b) and (f) show that $M_{\blambda,\bmu}$ grows linearly with $L=N$  for a simple scalable family of $\ket \bmu$ states \cite{comment_on_figure}.}
    \label{fig:stationary}
\end{figure}
Crucially, we find that the number of MCMC steps $\ell_{\rm max}$ needed to reconstruct $C(x,t)$ with high accuracy is drastically lower than the number of $\ket \bmu$ states that meaningfully contribute to the sum in \eqref{eq:Cxt2}. 
For thermal states this enables us to probe significantly higher temperatures and system sizes than previous approaches. 
We now focus on thermal expectation values, and set $h$ such that $n=N/L = 1$. We note that for $L \to \infty$ results for thermal states depend only on the two dimensionless parameters 
\be
\gamma=c\, n^{-1}\ ,\qquad \tau={T}\,{n^{-2}}\ .
\ee
In \cref{fig:stationary} we plot the values of $M_{\blambda,\bmu}/L$ sampled in single runs of MCMC, for several choices of $\tau$ and $\gamma$. As we increase $L=N$, the variance $\sigma_m^2$ of the distribution is seen to decrease, \emph{cf.}~also the inset of \cref{fig:s_vs_M_c_infty}, which shows for $\gamma=\infty$ and $\tau=20$ that $\sigma_m(L)\sim\alpha/L^\beta$ \footnote{We verified that a polynomial decay $\alpha/L^\beta$ fits well $\sigma_m(L)$ for all the pairs ($\tau$, $\gamma$) in \cref{fig:stationary}.}. This suggests that in the large $L$ limit the $\ket \bmu$ states that contribute non-negligibly to the Lehmann representation \eqref{eq:Cxt2} have $M_{\blambda,\bmu}/L$ values sharply peaked around an average $m[\rho]$ (here $\rho(\lambda)$ is the rapidity density, see Appendix \ref{appendix:GME}). 
\begin{figure}[t]
    \centering
    \includegraphics[scale=0.16]{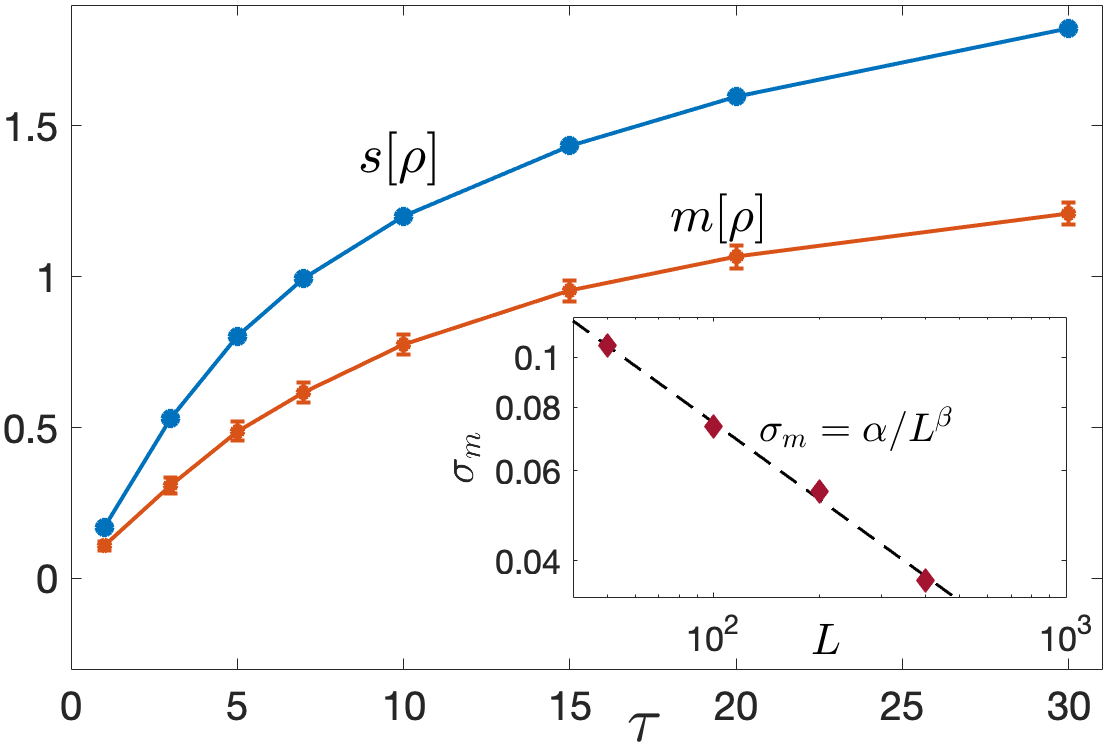}
    \caption{Average $m[\rho]=\braket{M_{\blambda,\bmu}}/L$ as a function of $\tau$ over single MCMC runs at $\gamma = \infty$ and $L=400$, \emph{cf}.~\cref{fig:stationary}(e). Blue symbols: entropy density $s[\rho]$ of the thermal macrostate at dimensionless temperature $\tau$, $\gamma = \infty$ and $L \to \infty$. The orange error bars indicate standard deviations $\sigma_m$. Inset: log-log plot for the standard deviation $\sigma_m$ as a function of $L$, for $\tau=20, \gamma=\infty$ (\emph{cf.}~\cref{fig:stationary}(e)). The 2-parameter linear fit (dashed line) yields $\alpha \approx 0.75$, $\beta\approx0.5$.}
    \label{fig:s_vs_M_c_infty}
\end{figure}
Given that the single-particle Green's function is finite in the thermodynamic limit, the corresponding average allows us to estimate the number of relevant states $|\bmu\rangle$ as $\sim e^{L m[\rho]}$. This should be compared to the total number of eigenstates in the thermal macrostate, which is $\sim e^{L s[\rho]}$, where $s[\rho]$ is the entropy density \cite{yang1969thermodynamics, korepin1993quantum_inverse} (see Appendix \ref{appendix:GME}). As evident from \cref{fig:stationary} (dashed lines), we find that $m[\rho]<s[\rho]$, indicating that only a vanishingly small fraction of thermal states (but still exponentially many in $L$) meaningfully contributes to the Lehmann sum, in agreement with the findings of Ref.~\cite{essler2024statistics}. 
We stress that $m[\rho]$ is a novel observable- and macrostate-dependent quantity that is a characteristic property of interacting integrable models. 
An explicit comparison of $m[\rho]$, estimated via single MCMC runs at $\gamma=\infty$, and $s[\rho]$ is shown in \cref{fig:s_vs_M_c_infty}.
The exponentially many states $\ket \bmu$ that matter in the Lehmann sum host an \emph{extensive} number of particle-hole excitations relative to the reference state $\ket \blambda$ \footnote{In fact, our numerical results imply that, in general, relevant $\ket{\bmu}$ states possess an extensive number of particle-hole excitations not only relative to $\ket{\blambda}$ ($N$-particle state), but also relative to any of the states $\ket{\bmu_{\rm ref}}$ ($N-1$-particle states) that are obtained by $N-1$ minimal-distance particle-hole excitations induced by changing the quantum numbers $I_j$ from half-odd integers to integers, \emph{cf.}~\cref{eq:logBetheEq} and \cite{essler2024statistics}.} (see also \cite{granet_finite_2020}). Yet, when compared with typical thermal states, they are characterized by sets of rapidities $\{ \mu_j\}$ that are anomalously close to $\{ \lambda_j \}$, as for these $M_{\bla,\bmu}\propto L$ (rather than being super-extensive) \cite{essler2024statistics}. {Examples of such linear (in $L$) scalings are given in the insets of \cref{fig:stationary}(b) and (f) for a specific family of $\ket \bmu$ states. The MCMC sampling provides a very efficient way of evaluating the Lehmann sum.
\begin{figure}[b]
\centering
\includegraphics[width=0.98\linewidth]{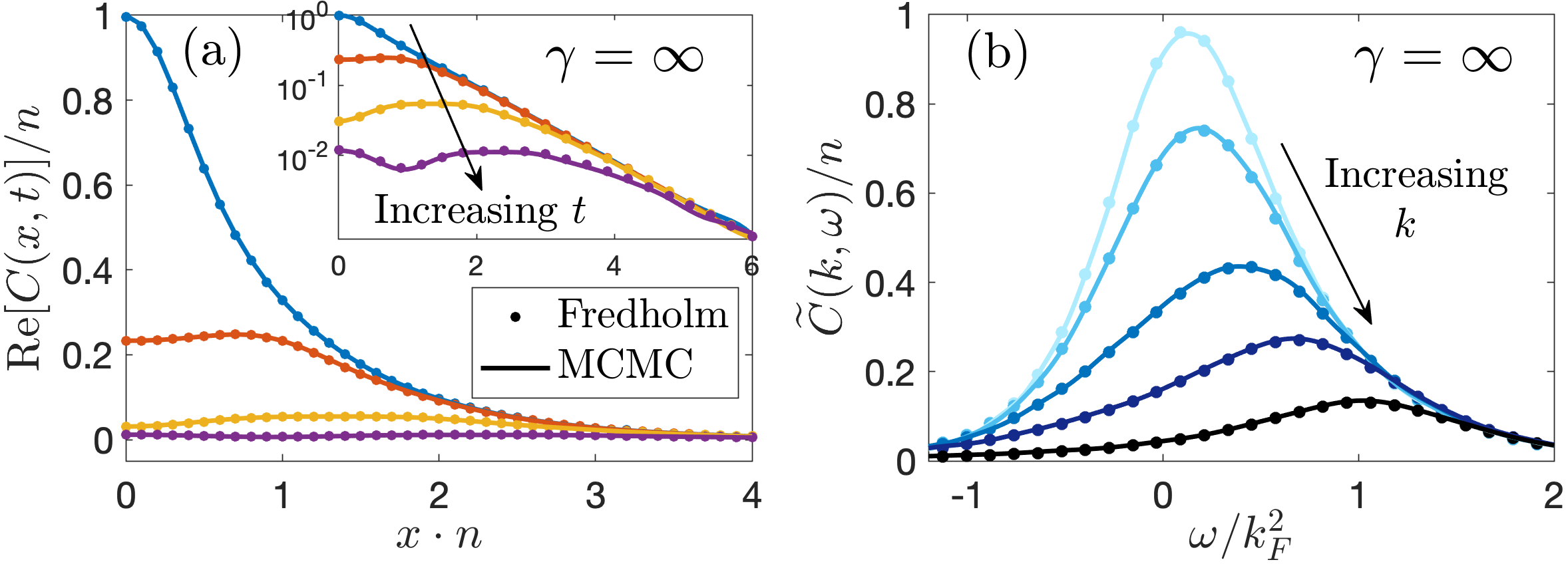}
\caption{$C(x,t)$ and $\tilde{C}(k,\omega)$ for $L=N = 200$, $\tau=5$ and $\gamma=\infty$, obtained by running 100 Markov chains in parallel with $\ell_{\rm max}=10^7$ MCMC steps each. The scale $k_F =\pi\, n$ is
the Fermi momentum at $\gamma=\infty$. (a) $\text{Re}[C(x,t)]$ as a function of position at times $t = (0, 0.2, 0.4, 0.6)/n^2$ (blue, red, yellow and purple curves). The dots are exact thermodynamic-limit results \cite{korepin1993quantum_inverse, korepin1990time_dependent}, see Appendix \ref{appendix:fredholm}. Inset: same data on log-scale. (b) $\tilde{C}(k,\omega)$ as a function of frequency at momenta $k/k_F = (0, 0.16, 0.32, 0.48, 0.8)$ (respectively from lightest to darkest curves) compared to exact results (dots).}
    \label{fig:benchmarks}
\end{figure}
For example, given that $m[\rho]\approx0.82$ and $\sigma_m\approx0.06$ at $\tau = 10, \gamma=10,L=200$ in \cref{fig:stationary}(d), we estimate that the number of relevant states is roughly $10^{66}$--$10^{76}$. This makes it impossible to apply algorithms like the ones used at $T=0$ and low $T$
\cite{caux2007one_particle, caux2009correlation, cheng_exact_2025,cheng2025one_body}. 
\begin{figure}[b]
\centering
\includegraphics[width=0.98\linewidth]{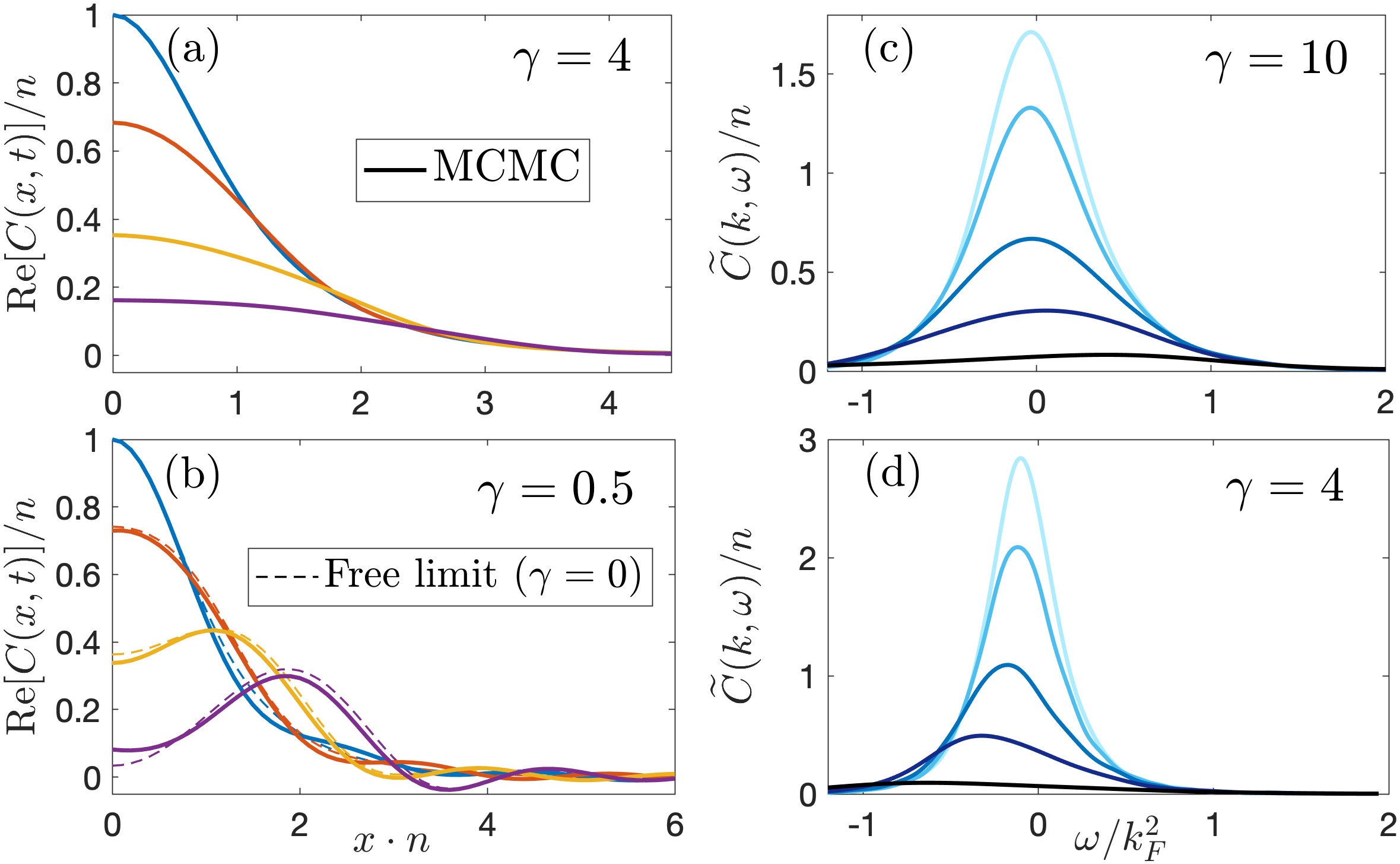}
\caption{Same $\tau=5, L =200$ MCMC plots as in \cref{fig:benchmarks}, but for finite values of $\gamma = 0.5, 4, 10$. (a), (b): $\text{Re}[C(x,t)]$ as a function of position at times $t = (0, 0.2, 0.4, 0.6)/n^2$ (blue, red, yellow and purple curves). (c), (d): $\tilde{C}(k,\omega)$ as a function of frequency at momenta $k/k_F = (0, 0.16, 0.32, 0.48, 0.8)$ (lightest to darkest curves). $\ell_{\rm max}=10^6$ in each of the 100 Markov chain runs. Dashed lines in (b) are exact $L\to\infty$ results in absence of interactions ($\gamma=0$).}
    \label{fig:finite_gamma}
\end{figure}
However, we will now demonstrate that very accurate results for $C(x,t)$ can be obtained using MCMC sampling techniques that require a total of only $10^8$--$10^9$ steps. To extract information on the spectrum of physical excitations created by the action of $\phi(x)$ at finite temperatures it is useful to consider the Fourier transform of $C(x,t)$ \cite{caux2007one_particle}
\begin{equation}
\begin{aligned}
    &\widetilde{C}(k,\omega)=\int_{0}^{L}dx \int_{-\infty}^\infty dt \, e^{i \omega t - i k x}C(x,t) \\
    & \ \ \ = 2 \pi L \sum_{\bmu} \delta(\omega-E_{\bmu} + E_{\blambda}) \, \delta_{k,P_{\bmu}-P_\blambda} \,  |\braket{\bmu | \phi(0,0)|\blambda}|^2 \ .
    \end{aligned}
\end{equation}
While it is in principle possible to compute $\widetilde{C}(k,\omega)$ by Fourier transforming the MCMC data for $C(x,t)$, it is more convenient to sample directly in Fourier space. Given that at any finite $L$ the function $\widetilde{C}(k,\omega)$ consists of a sum of Dirac delta peaks, we implement a smoothing procedure appropriate for comparison with $L \to \infty$, as detailed in Appendix \ref{appendix:reconstructingtildeC}.

%
%
\begin{figure*}[t]
    \centering
    \includegraphics[width=0.9\linewidth]{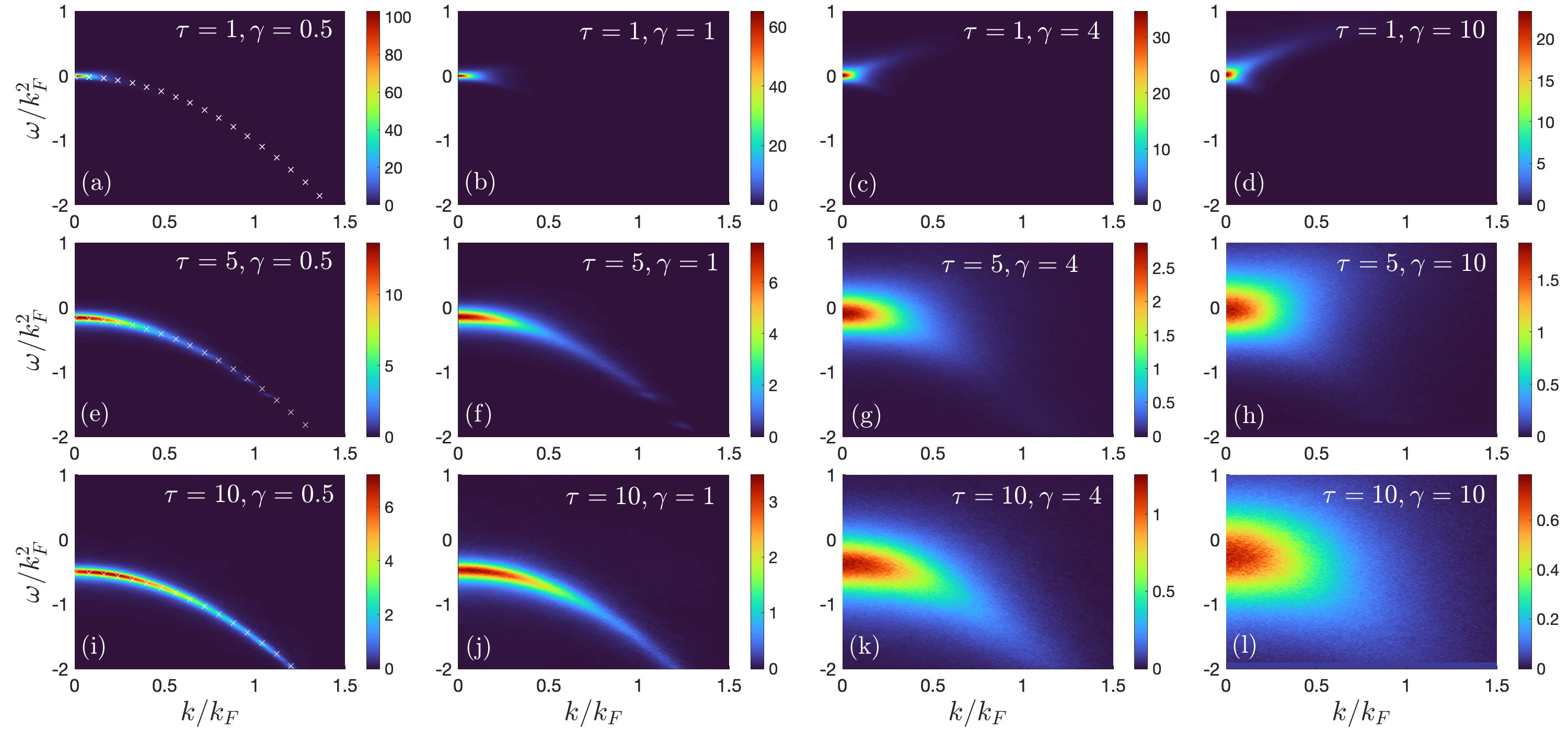}
    \caption{Density plots for $\widetilde{C}(k,\omega)/n$ at several values of $\tau$ and $\gamma$. Data obtained for $L=200$ by running 100 parallel Markov chains with $\ell_{\rm max}=10^6$ steps each. The white ‘x’ markers in (a), (e), (i) denote the continuous free dispersion $\omega=-k^2+h_{\gamma=0}$, where $h_{\gamma=0}$ is the chemical potential that sets $n=1$ at $\gamma=0$. Note the different scales of colours for each subplot.}
    \label{fig:final_density_plots}
\end{figure*}
\section{Benchmarks and results}
The $c\to\infty$ limit of the single-particle Green's function is non-trivial, because the Bose field is non-local with respect to the free fermion modes that diagonalize the Hamiltonian \cite{creamer1980quantum}. As a result,
the form factors  $\braket{\bmu|\phi(0,0)|\blambda}$ are non-vanishing for any  $\ket \bmu$ with $N-1$ particles \cite{caux2007one_particle}, just like for $\gamma < \infty$ (\emph{cf}. Appendix \ref{appendix:formfactors}). Therefore, the limit $\gamma = \infty$ poses the same challenges for MCMC as the generic case $\gamma<\infty$, while at the same time
analytical expressions for $\lim_{L \to \infty}C(x,t)\big|_{\gamma=\infty}$ exist \cite{korepin1993quantum_inverse, korepin1990time_dependent, lenard1964momentum, lenard1966one_dimensional, patu2022exact} and enable a direct comparison with exact results. The analytic results involve Fredholm determinants that can be numerically evaluated by standard methods \cite{bornemann2010numerical}, see Appendix \ref{appendix:fredholm}.

In \cref{fig:benchmarks}(a) and (b) we show MCMC results for $\text{Re}[C(x,t)]$ and
$\widetilde{C}(k,\omega)$ at $\gamma=\infty$, $\tau=5$ and $L=200$. The agreement with the exact results is clearly excellent (the same is true for $\text{Im}[C(x,t)]$). 
We note that the MCMC results correctly describe the exponential decay with $x$ expected at finite temperatures, \emph{cf.}~inset of \cref{fig:benchmarks}(a). In the static limit ($t=0$) we can benchmark our results against 
Ref.~\cite{patu_correlation_2013} for arbitrary temperatures and $0<\gamma<\infty$, see Appendix \ref{appendix:staticbench}. Additional benchmarks for $t>0$ at (i) $\gamma=\infty$ and general $\tau$  and (ii)
general $\gamma>0$ for $\tau=0$ are given in Appendix \ref{appendix:additionalbenchandres}.\\

In \cref{fig:finite_gamma}(b) we compare MCMC data at $\gamma=0.5$ to the exact thermodynamic-limit results in the absence of contact interactions. 
The two sets of results are seen to be very close to one another, in agreement with the expectation that
small $\gamma$ should tend towards the free limit $\gamma=0$. \cref{fig:finite_gamma}(a) shows results for the intermediate value of $\gamma=4$. In \cref{fig:finite_gamma}(c) and (d) we show Fourier space results for $\gamma = 4,10$. We see that at fixed $\tau>0$ a decrease in $\gamma$ causes the curves to become more sharply peaked and to drift towards negative values of $\omega$, \emph{cf.}~also \cref{fig:benchmarks}(b). The former behaviour is consistent with the emergence, at low interaction couplings, of well-defined quasiparticles that couple to $\phi(x)$ and whose damping rate vanishes as the non-interacting limit $\gamma=0$ is approached. Similarly, the drift towards negative energies can be understood from the fact that at $\gamma=0$ the Fourier correlator is a sum of delta functions following the free dispersion $\omega(k) = -k^2 + h$ at minus the occupied momenta of $\ket \blambda$, i.e~at $k_i=-\lambda_i$.
In \cref{fig:final_density_plots} we show density plots for $\widetilde{C}(k,\omega)$ at several temperatures $\tau$ and couplings $\gamma$. For $\gamma=0.5$ the spectral weight closely follows the free dispersion $\omega = -k^2 + h$ of the non-interacting limit $\gamma=0$. In fact, for $\gamma\le1$ we observe a fairly sharp single-particle mode whose peak is slightly broadened by temperature and interactions. At higher $\gamma$ (especially for $\tau \ge 5$) the spectral weight spreads out into a continuum. The plots at $\tau=1$ and high $\gamma$ are consistent with the zero and very-low-temperature results of \cite{caux2007one_particle, cheng2025one_body}, with a broadened “type II” dispersion \cite{lieb1963exact_2} visible for $\omega>0$ (see Appendix \ref{appendix:additionalbenchandres} for additional density plots in log-scale, which better showcase this feature). 
\begin{figure}[b]
    \centering
    \includegraphics[width=0.6\linewidth]{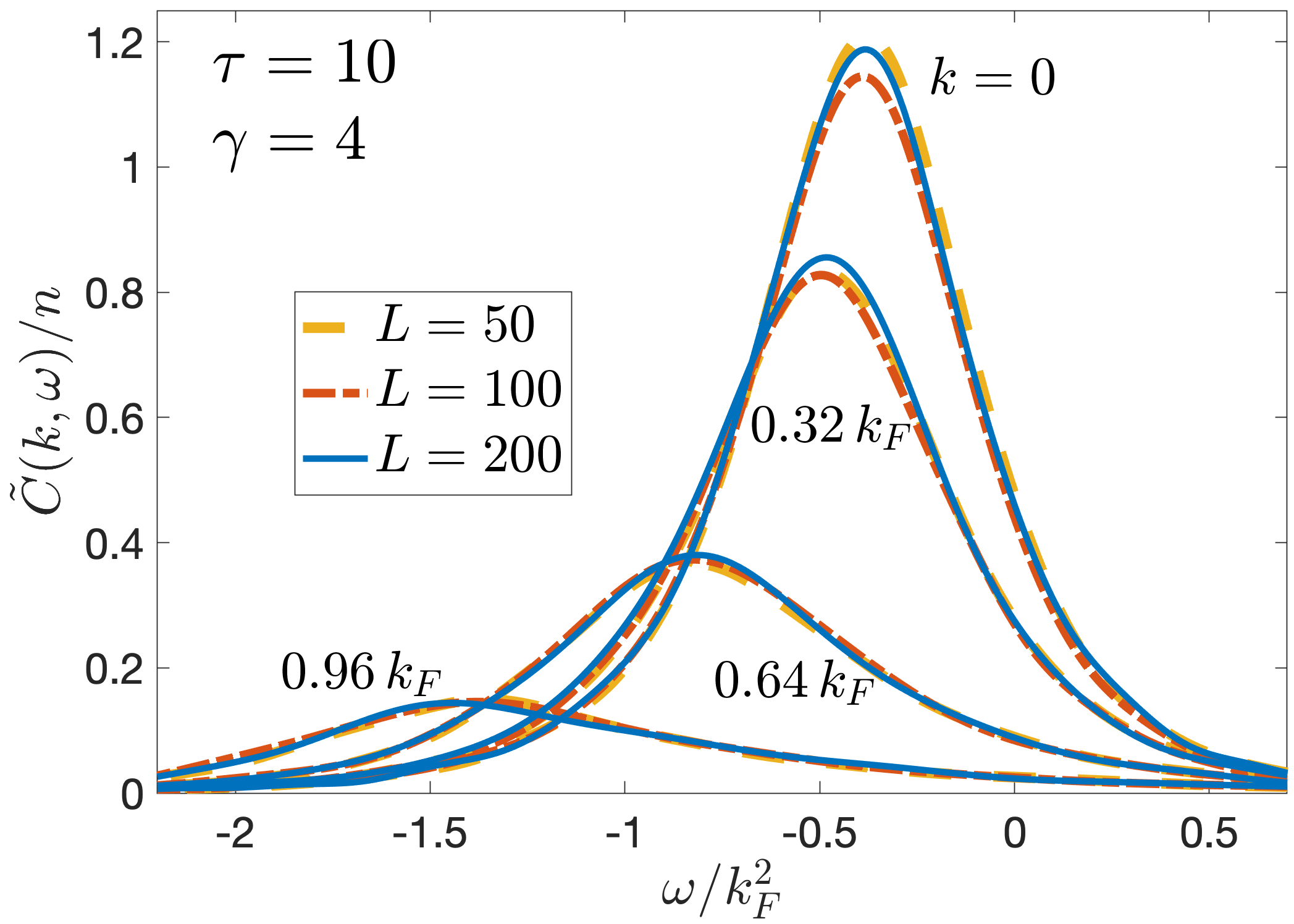}
    \caption{Finite size effects for $\widetilde{C}(k,\omega)$, at $\tau=10$ and $\gamma=4$. The frequency and momenta are rescaled with $k_F=\pi \, n$.}
    \label{fig:finite_size_effect}
\end{figure}
An important question is how close the system sizes we can reach in practice are to the thermodynamic limit. To address this issue we show $\widetilde{C}(k,\omega)$ for several values of $L$ in \cref{fig:finite_size_effect}. We see that in the range of $k$ and $\omega$ values of interest (where the signal is significant) finite size effects are very small already at $L\gtrsim 50$, given the good agreement of all the curves (see also additional plots in Appendix \ref{appendix:additionalbenchandres}). 
In \cref{fig:GGE} we show $\widetilde{C}(k,\omega)$ in GGEs (\emph{cf}.~Appendix \hyperlink{page.8}{A}). By choosing $\beta_3 \neq 0$ we break the parity, which leads to marked differences compared to the thermal case of \cref{fig:final_density_plots} at the same values of $\gamma$ and $T=1/\beta_2$ (note that also here we set $n=1$). The rapidity densities $\rho(\lambda)$ of the GGE and thermal states used, as well as GGE results for $\gamma=1$, are shown in Appendix \ref{appendix:additionalbenchandres}.

\begin{figure}[t]
    \centering
    \includegraphics[width=0.95\linewidth]{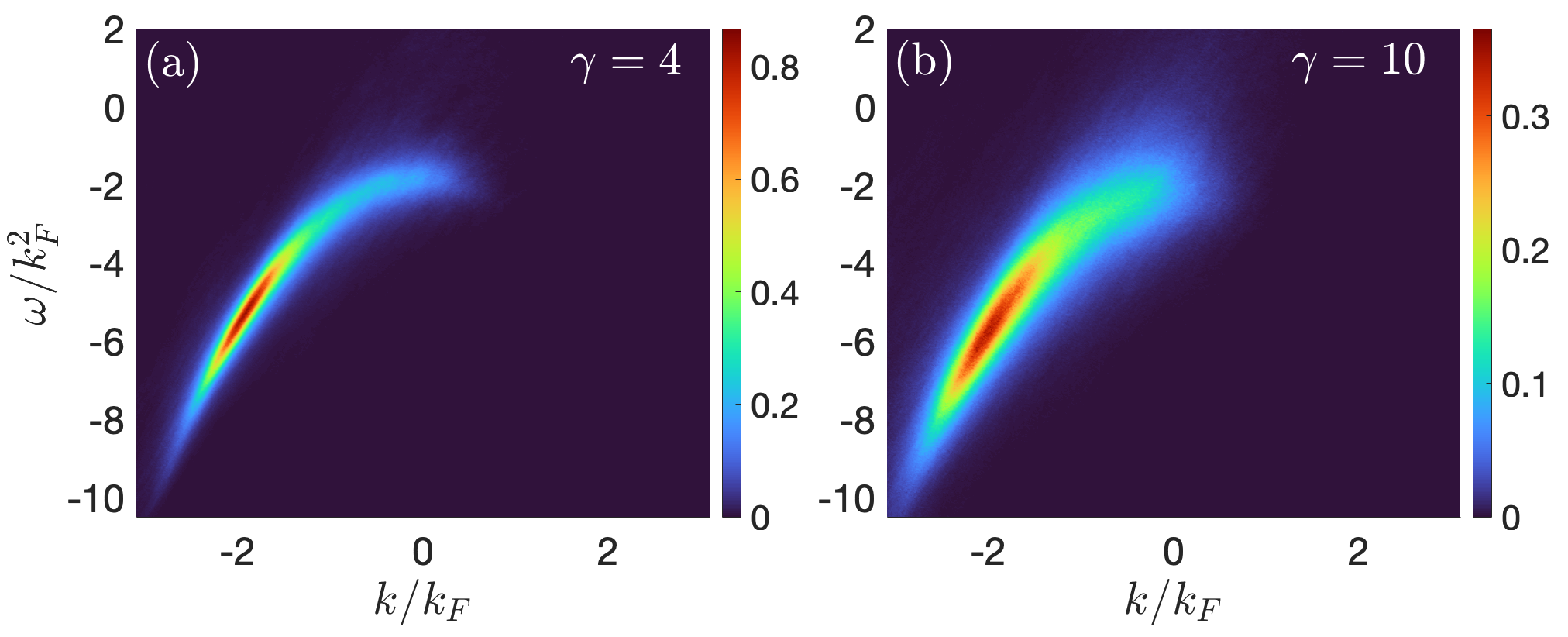}
    \caption{Density plots for $\widetilde{C}(k,\omega)/n$ where $\ket \bla$ belongs to a GGE macrostate (\emph{cf}.~Appendix \ref{appendix:GME}). (a) At $\gamma=4$, the GGE is characterized by $T = 1/\beta_2=10$, $h = -\beta_0/\beta_2 =-9.2$, $\beta_3 = -0.042$, $\beta_4 = 0.0035$ and $\beta_j = 0$ for $j =1$ and $j\ge 5$. (b) At $\gamma=10$: $T=10$, $h = -6.83$, $\beta_3=-0.036$, $\beta_4 = 0.0027$ and $\beta_j=0$ for all other $j$. As before, the value of $h$ is chosen to fix $n=N/L=1$. Both plots are for system size $L=200$.}
    \label{fig:GGE}
\end{figure}
%
%

\section{Discussion}
We have developed a Monte Carlo sampling method for evaluating dynamical correlation functions in the LL model and applied it to the two-point function of the Bose field. We have presented results in a large range of temperatures and interaction strengths for thermal ensembles, and also considered the dynamics in GGEs. We have shown that an exponentially large (in $L$) number of form factors needs to be sampled, which is outside the range of applicability of approaches originally developed for $T=0$ dynamics \cite{caux2007one_particle,caux2009correlation,de_klerk2023improved,cheng2025one_body}. Interestingly, these methods have been successfully applied to $T>0$ dynamical density-density correlation functions for fairly large values of $N \approx 50$--$100$ \cite{panfil2014finite}. This is possible because the structure of form factors of $\varrho(x)=\phi^\dag(x) \phi(x)$ in that range of $N$ and $L$ does not yet require sampling a prohibitively large number of terms, see Appendix \ref{eq:densityvsBosefield} for a more detailed discussion. Our work opens the door to using MCMC sampling of Lehmann representations in other contexts, in particular quantum quenches \cite{SeneseEsslerUnpublished}\cite{caux2013time,robinson2021computing}.

\begin{acknowledgments}
\textit{Acknowledgments.---}%
This work was supported in part by the EPSRC under grant EP/X030881/1. We thank B. Bertini for helpful discussions.
\end{acknowledgments}


\bibliography{apssamp}

\begin{thebibliography}{94}
\bibitem [{com()}]{comment_on_figure}%
  \BibitemOpen
  \href@noop {} {}\bibinfo {note} {In particular, we plot the average $M_{\bla,\bmu}$ for $100$ eigenstates $\ket \bmu$ randomly generated by solving the Bethe equations \eqref{eq:logBetheEq} with integers $\{J^{(\bmu)}_j=\tilde{I}_j^{(\bla)}\pm 1/2\}_{j=1}^{N-1}$ (the signs $\pm$ are drawn at random), where $\{\tilde{I}_j^{(\bla)}\}$ are $N-1$ of the $N$ half-odd integers $I_j^{(\bla)}$ that the determine the rapidities $\lambda_j$.}\BibitemShut {Stop}%
\end{thebibliography}

\clearpage
\onecolumngrid

\appendix

\section{Generalized microcanonical ensemble}
\label{appendix:GME}

In the limit of large $L$ at fixed $n=N/L$, physical properties of the LL model are conveniently described in terms of \emph{macrostates} \cite{yang1969thermodynamics, korepin1993quantum_inverse, takahashi1999thermodynamics}. These are families of eigenstates with the same  “root density” (or “rapidity” density) $\rho(\lambda)$ \footnote{In this work we consider only macrostates characterized by root densities that vanish faster than any power of $1/\lambda$ when $\lambda\to\infty$.}, which defines their rapidity distributions as
\begin{equation}
\big\lfloor L \rho(\lambda) d\lambda\big\rfloor = \text{number of } \lambda_j \text{\ in} \ [\lambda, \lambda + d \lambda] \ .
\end{equation}
By construction eigenstates belonging to the same macrostate have, up to finite-size corrections that vanish in the thermodynamic limit, the same local properties \cite{korepin1993quantum_inverse,essler2024statistics}. We are interested in macrostates corresponding to GGEs
\begin{align}
\langle {\cal O} \rangle_{\rm GGE}=\frac{1}{Z_{\rm GGE}}{\rm Tr}\left[{\cal O} \, e^{-\sum_{n\geq 0}\beta_nQ^{(n)}}\right].
\label{GGE}
\end{align}
By “corresponding” we mean that such GGE expectation values are dominated (exponentially with $L$) by eigenstates belonging to the macrostate $\rho(\lambda)$ of highest entropy density $s[\rho]$ under the constraints $\int d \lambda\, \rho(\lambda) \lambda^n = \braket{Q^{(n)}}_{\rm GGE}/L$, where $s[\rho]$ is defined by
\begin{equation}
\begin{aligned}
    s[\rho]&=\int_{-\infty}^{\infty}d\lambda \, [\rho_t(\lambda)\log \rho_t(\lambda)-\rho(\lambda)\log \rho(\lambda)-\rho_h(\lambda)\log \rho_h(\lambda)] \ ,\\
    \rho_t(\lambda)&=\frac{1}{2\pi}\left[1+\int_{-\infty}^\infty d \mu \, K(\lambda-\mu)\rho(\mu)\right] \quad\qquad \qquad \rho_h(\lambda) = \rho_t(\lambda) - \rho(\lambda) \ ,
\end{aligned}
\end{equation}
and $K(\lambda)= \frac{2c}{c^2+\lambda^2}$. 
The thermal (Gibbs) ensemble is obtained from \eqref{GGE} by the choice $\beta_0=-h/T$, $\beta_2=1/T$ and all other $\beta_j=0$. The macrostate root density corresponding to (\ref{GGE}) is obtained from the integral equations \cite{yang1969thermodynamics}
\begin{align}
&\epsilon(\lambda)=\sum_{n\geq 0}\beta_n\lambda^n-\int_{-\infty}^\infty \frac{d\mu}{2\pi}\ K(\lambda-\mu)\ln\big(1+e^{-\epsilon(\mu)}\big),\nn
&\rho(\lambda)\big(1+e^{\epsilon(\lambda)}\big)=\frac{1}{2\pi}+\int_{-\infty}^\infty \frac{d\mu}{2\pi}\ K(\lambda-\mu)\rho(\mu)\ .
\end{align}
In practice, it is convenient to work with the generalized microcanonical ensemble, which replaces \eqref{GGE} by the expectation value on a single ($N$-particle) representative eigenstate (microstate) $\ket{\blambda}$ corresponding to $\rho(\lambda)$ \cite{korepin1993quantum_inverse,caux2013time,essler2016quench}. In particular, we construct $\ket \blambda$ by requiring the finite-$N$ distribution of its Bethe rapidities to smoothly match the profile of the thermodynamic root density $\rho(\lambda)$ chosen, see e.g.~\cite{essler2024statistics}.  The replacement of \eqref{GGE} with the expectation value on $\ket{\bla}$ leads to differences that vanish as $L \to \infty$ \cite{korepin1993quantum_inverse}.\\

\section{Details of Markov chain Monte Carlo sampling}
\label{appendix:detailsofMCMC}

For completeness, we start by reviewing a few elementary results from the theory of Markov processes \cite{feller1957introduction} that are relevant for MCMC sampling \cite{rubinstein2016simulation}. Consider a finite set $\mathcal{E}=(y_1, y_2,\ldots, y_R)$ and a stochastic variable $Y^{(\ell)}$ that takes values in $\mathcal{E}$ at discrete steps $\ell$. Consider a Markov chain on $\mathcal{E}$ with transition matrix $T_{ij}$, defined as the probability of transitioning from $y_i$ to $y_j$ in one time step
\begin{equation}
\label{eq:PtransMCMC}
    \mathbb{P}(Y^{(\ell+1)}=y_j | Y^{(\ell)}=y_i) = T_{ij} \ ,
\end{equation}
where $\mathbb{P}(\cdot | \cdot)$ indicates conditional probability. If we call $\pi^{(\ell)}_i$ the probability associated with $Y^{(\ell)}=y_i$, we have 
\begin{equation}
    \pi^{(\ell)} = \pi^{(0)} T^\ell \ \qquad \qquad \quad \pi^{(\ell)} = (\pi_1^{(\ell)}, \pi^{(\ell)}_2, \ldots, \pi_R^{(\ell)})\ .
\end{equation}
Consider transitions matrices $T$ that are irreducible and aperiodic \cite{feller1957introduction}. Then, there exists a limiting distribution $\lim_{\ell\to\infty}\pi^{(\ell)}\equiv \pi$ given by the unique solution of
\begin{equation}
\label{eq:stationarityTmatrix}
    \pi = \pi \, T \ .
\end{equation}
From this we see that a simple way of constructing a Markov chain with stationary distribution $\pi$ is to require that $T$ satisfies the \emph{detailed balance} condition 
\begin{equation}
\label{eq:detailedbalance}
    \pi_i T_{ij} = \pi_j T_{ji} \ .
\end{equation}
Indeed, \eqref{eq:stationarityTmatrix} follows from \eqref{eq:detailedbalance} by simply noticing that for every transition matrix $\sum_{j}T_{ij}=1$. In our context, $\pi$ represents the target probability distribution $\mathcal{P}_\blambda(\bmu)=\exp(-M_{\blambda,\bmu})/Z_{\blambda}$ appearing in \cref{eq:Cxt2}. 
To construct a $T$ that satisfies detailed balance one can employ the Metropolis-Hastings algorithm \cite{metropolis1953equation, hastings1970monte, rubinstein2016simulation}. This allows an approximate generation of samples $Y^{(\ell)}$ from the target distribution $\pi$, by letting $\ell$ run up to large values. 
The Metropolis-Hastings construction is based on a proposal matrix $w_{ij}$, which denotes the probability of proposing the move $Y^{(\ell)}=y_i \to Y^{(\ell+1)}=y_j$, and on accepting or refusing it with probability, respectively, $\alpha_{ij}$ and $1-\alpha_{ij}$. This is defined as
\begin{equation}
\label{eq:MetropolisAcceptance}
    \alpha_{ij} = \min \left[ \frac{\pi_j\, w_{ji}}{\pi_i \, w_{ij}} , 1 \right] \ .
\end{equation}
It is easy to see that the overall transition matrix $T$ given by
\begin{equation}
    T_{ij} = w_{ij} \, \alpha_{ij} \quad i \neq j \qquad \qquad \qquad T_{ii} = 1 - \sum_{\substack{j \\ (j \neq i)}} w_{ij}\alpha_{ij} 
\end{equation}
satisfies the detailed balance of \cref{eq:detailedbalance}. This implies that, if we choose $w_{ij}$ in such a way that $T$ is irreducible and aperiodic, then at large $\ell$ steps $Y^{(\ell)}$ will be sampled according to $\pi$, irrespective of the initial value $Y^{(0)}$.\\

For computing dynamical correlation functions in the LL model, we aim to sample $\ket \bmu$ eigenstates according to the probability distribution $\mathcal{P}_\blambda(\bmu)$, where $\ket \blambda$ is a fixed $N$-particle eigenstate (and we assume $N$ to be even). Given that $\ket \bmu$ eigenstates with $N-1$ particles are in one-to-one correspondence with the sets of $N-1$ distinct integers $\{J_j\}_{j=1}^{N-1}$ that enter the Bethe equations (see \cref{eq:logBetheEq}), we can directly perform sampling over the space of integers. This means that each element $y_i$ entering \eqref{eq:PtransMCMC} is now a set of $N-1$ distinct integers $\{J_j\}$, and we guarantee finiteness of $\mathcal{E}$ by imposing any arbitrarily-large cutoff $J_{\rm max}$ on $\max |J_j|$. The intermediate numerical solution of the Bethe equations \eqref{eq:logBetheEq} needed at each step can be performed very efficiently by using the Newton-Raphson method. The proposal matrix $w$ is implicitly defined by the following \emph{single-integer} proposal scheme. For the step $\ell \to \ell +1$ we select at random one among the $N-1$ integers $J_j^{(\ell)}$, say $j=s$. We then consider the nearest $Q/2$ \emph{unoccupied} integers both to its right and left ($Q$ even), and select one at random, call it $J^*$ \footnote{If we hit the boundaries defined by $J_{\rm max}$ we wrap around, but in practice $J_{\rm max}$ can be chosen large enough that this never happens.}. The matrix $w_{ik}$ associated with the proposed update $J_{s}^{(\ell)}\to J^*$ is symmetric, i.e. $w_{ik}=w_{ki}$. Once the move has been proposed, we accept it according to the Metropolis probability
\begin{equation}
\label{eq:MetropolisAcceptanceAppendix}
    \alpha_{ik} = \min \left[ \frac{P_\blambda(\bmu_k)}{P_\blambda(\bmu_i)} , 1 \right] \ ,
\end{equation}
where $\ket{\bmu_i}$ is the eigenstate whose quantum numbers host the integer $J_s^{(\ell)}$ and $\ket{\bmu_k}$ the one that hosts $J^*$ in its place. We note that $\alpha_{ik}$ in \eqref{eq:MetropolisAcceptanceAppendix} does not depend on $w_{ik}$ as a consequence of the symmetry $w = w^T$, as in the original algorithm of \cite{metropolis1953equation}. If the move is accepted we update $J_s^{(\ell+1)}=J^*$, otherwise $J_s^{(\ell+1)}=J_s^{(\ell)}$. It is also possible to update $m$ integers at the time by iterating $m$ times the single-integer proposal defined by $w_{ik}$. Given that $(w^m)_{ik}$ remains symmetric, also in this case $\alpha_{ik}$ depends only on $\mathcal{P}_\blambda(\bmu)$. We note that our choice of $w_{ik}$ yields an overall transition matrix for the Markov process which is irreducible and aperiodic \cite{feller1957introduction}, and this, together with \eqref{eq:MetropolisAcceptanceAppendix}, guarantees that in the limit of large $\ell$ the probability of having $Y^{(\ell)}$ equal to a given set of integers $y_i \ (\leftrightarrow \ket{\bmu_i})$ converges to $\mathcal{P}_\blambda(\bmu_i)$, irrespective of the chosen initial set $Y^{(0)}$. 
\begin{figure*}[th]
    \centering
    \includegraphics[width=0.85\linewidth]{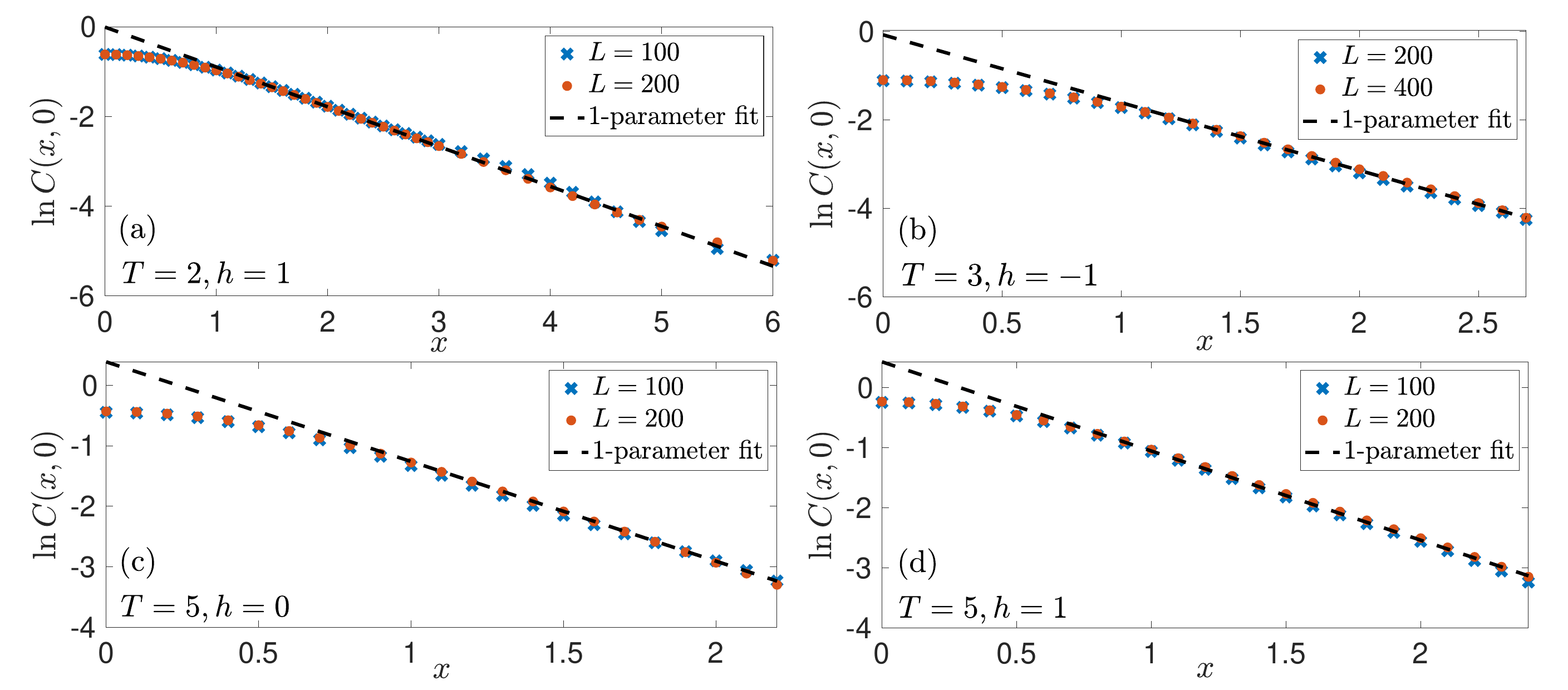}
    \caption{Decay of $\ln C(x,0)$ and 1-parameter linear fit over $\delta$ (see \eqref{eq:fitfunction}), assuming $\alpha = 1/\xi_{\rm PK}$ and $\sigma=0$. The interaction coupling is equal to $c=2$ in all subplots. Unlike the main text (where we fix $n=1$), the density $n$ is fixed by $T$ and $h$ chosen to compare with \cite{patu_correlation_2013} and is equal to (respectively for subplots a, b, c and d) $n \approx$ 0.542, 0.339, 0.651, 0.792. The MCMC data have been obtained using 100 parallel Markov chains, each run up to $10^7$ steps. The fits are performed over the larger-$L$ dataset.}
    \label{fig:corr_length}
\end{figure*}
For the data presented in the main text, we have
\begin{enumerate}
    \item Tuned the values of $Q$ and $m$ such to always have an average acceptance rate of at least $\approx10\%$. We find that setting $m=1$ and adjusting $Q/2$ in the range $[1, 8]$ is always sufficient to obtain accurate sampling of the stationary distribution $P_\blambda(\bmu)$. 
    \item Verified that irrespective of the initial set $\{J_j^{(0)}\}$ chosen, the Markov chain always ends up sampling from the same distribution (which appears stationary) at large-enough number of steps $\ell$. This was achieved by monitoring the values of $M_{\blambda,\bmu}$ as in \cref{fig:stationary}, or those for the energies $E_\bmu$ and momenta $P_\bmu$. 
    \item Discarded a “burn-in” number of initial steps $\ell$, to allow time for the Markov chain to reach the stationary distribution. We set $\ell_{\rm burn-in}$ to be $1/11$ of the total number of MCMC steps $\ell_{\rm burn-in}+\ell_{\rm max}$. 
    \item Extracted the statistical uncertainty on the final MCMC data by computing the standard error on the mean associated with the output of $C$ Markov chains run in parallel, where usually we set $C=100$. We note that it is also possible to estimate uncertainties using the “batch” method \cite{gilks1995markov} within a single, possibly longer, Markov chain run. The uncertainties obtained were negligible in all regimes of physical interest. 
\end{enumerate}

\section{Benchmark for correlation length at finite interaction strength}
\label{appendix:staticbench}
As mentioned in the introduction, there are no analytic results available for dynamical correlators in the interacting regime $0< c <\infty$ of the LL model that enable a direct benchmark of our method. However, restricting the attention to the static case $(t=0)$, Pâtu and Klümper (PK) managed to obtain an expression (that one can numerically evaluate) for the asymptotic correlation length $\xi$ associated with the large distance decay of $C(x,0) \in \mathbb{R}$ from \cref{eq:Cxt}, at arbitrary temperatures and interaction strengths $c$ \cite{patu_correlation_2013}. 
Therefore, we can compare the correlation length $\xi$ extracted from our MCMC results with $\xi_{\rm PK}$ reported in Fig.~3 of \cite{patu_correlation_2013} (where the interaction strength is fixed to $c=2$). The correlation length $\xi$ can be extracted from the large-$x$ decay by a 3-parameter fit
\begin{equation}
\label{eq:fitfunction}
    \ln C(x,0) = - \alpha \, x - \sigma \ln x + \delta \qquad \quad \xi \ \ \to \ \ 1/\alpha \ ,
\end{equation}
where $\alpha>0$, $\sigma>0$. In all the preliminary fits performed, the value of $\sigma$ was always found to be very close to zero, suggesting that there are no polynomial corrections to the exponential decay. This is in general not surprising in 1D, and matches the expected asymptotic form \cite{patu_correlation_2013}.

To verify the agreement between the Monte Carlo data and the correlation length $\xi_{\rm PK}$ from \cite{patu_correlation_2013}, in \cref{fig:corr_length} we perform \footnote{The range of $x$ values considered in the fit is restricted to the intermediate region that clearly showcases linear behaviour. In other words, we discard the nonlinear features at small $x$ and the noisy regions (due to statistical uncertainties from MCMC) at large $x$.} a \emph{1-parameter} fit over $\delta$ from \eqref{eq:fitfunction}, while setting $\alpha = 1/\xi_{\rm PK}$ and $\sigma = 0$. The agreement is excellent for the 4 examples shown. 
A second check is performed without assuming a priori knowledge of $\xi_{\rm PK}$ from \cite{patu_correlation_2013}. We restrict to the intermediate-$x$ region which showcases a (almost) perfect linear trend, and we perform a 2-parameter fit over $\alpha$ and $\delta$ from \eqref{eq:fitfunction} ($\sigma = 0$). In \cref{fig:thermal_comparison} we compare the values of $\xi=1/\alpha$ obtained in this way (and their uncertainties) with those from Fig.~3 of \cite{patu_correlation_2013}. The agreement is very good. 

\begin{figure}[t!]
    \centering
    \includegraphics[width=0.43\linewidth]{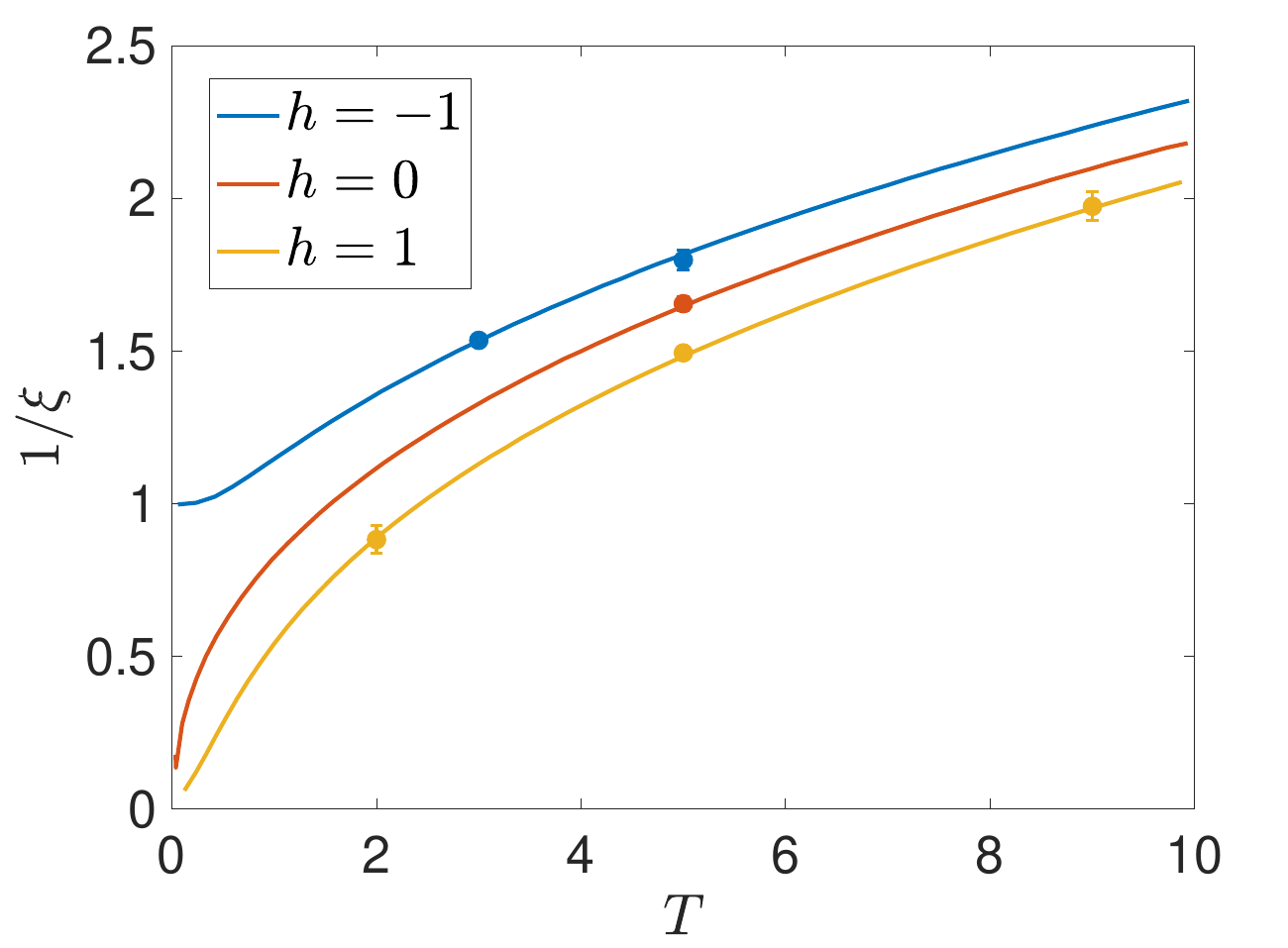}
    \caption{For $c=2$, comparison of the values of $1/\xi_{\rm PK}$ from Fig.~3 of \cite{patu_correlation_2013} (lines) with those extracted from Monte Carlo data (dots) via 2-parameter linear fits over $\alpha$ and $\delta$ in \eqref{eq:fitfunction} ($\sigma = 0$). The error bars indicate $95\%$ confidence intervals.}
    \label{fig:thermal_comparison}
\end{figure}

\section{Reconstructing $\widetilde{C}(k,\omega)$}
\label{appendix:reconstructingtildeC}

The correlation function in Fourier space is 
\begin{equation}
    \widetilde{C}(k,\omega)= 2 \pi L \sum_\bmu \delta(\omega-E_\bmu + E_\blambda) \, \delta_{k,P_\bmu-P_\blambda} \,  |\braket{\bmu | \phi(0,0)|\blambda}|^2  \ .
\end{equation}
At any finite $L$ it consists of a sum of Dirac delta peaks centred around the energy difference $E_\bmu - E_\blambda$. Due to this, it is more convenient to sample an integral of $\widetilde{C}(k,\omega)$ over the frequency
\begin{equation}
    \widetilde{C}_{\rm int}(k,\omega) = \int_{-\infty}^\omega d y\, \widetilde{C}(k,y) = 2 \pi L \sum_\bmu \theta_H(\omega -E_\bmu + E_\blambda) \, \delta_{k,P_\bmu-P_\blambda} \,  |\braket{\bmu | \phi(0,0)|\blambda}|^2  \ ,
\end{equation}
where $\theta_H$ is the Heaviside step function. Fixed $k$, the curve $\widetilde{C}_{\rm int}(k,\omega)$ as a function of $\omega$ consists of many discrete steps of height $\propto |\braket{\bmu | \phi(0,0)|\blambda}|^2=e^{-M_{\blambda,\bmu}}$ separated by a distance $\delta \omega \sim D_\bmu(E)^{-1}$, where $D_\bmu(E)$ indicates the density of $\ket \bmu$ states in energy (which grows exponentially with $L$ at any finite energy density). However, if one resolves the curve $\widetilde{C}_{\rm int}(k,\omega)$ ($k$ fixed) only on a finite grid of $\omega$ values with $\Delta \omega_{\rm grid}\gg \delta \omega$, it appears to be smooth, enabling a direct comparison with the smooth function $\lim_{L\to\infty}\widetilde{C}_{\rm int}(k,\omega)$ in the thermodynamic limit. 

\begin{figure}[b]
    \centering
    \includegraphics[scale=0.19]{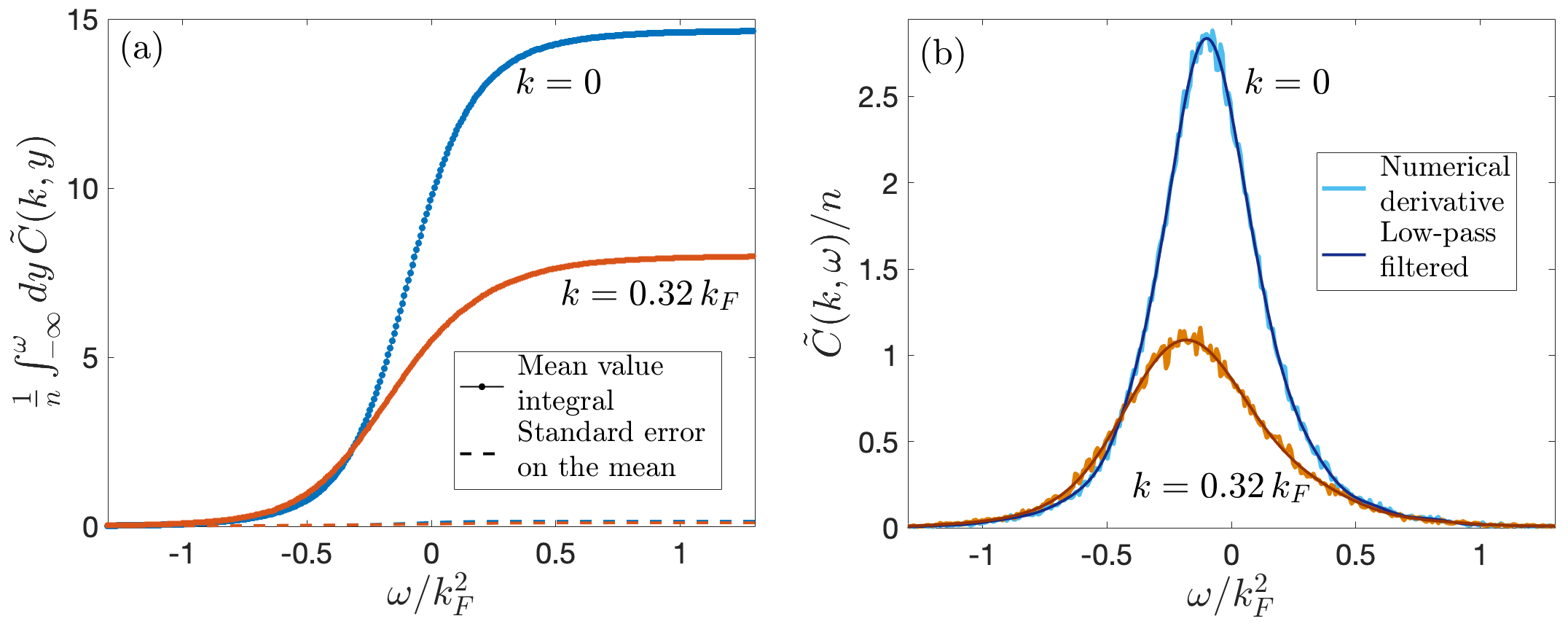}
    \caption{(a) $\widetilde{C}_{\rm int}(k,\omega)=\int_{-\infty}^\omega dy\,\widetilde{C}(k,y)$ obtained by running 100 parallel Markov chains with $\ell_{\rm max}=10^6$ MCMC steps each at $\tau=5, \gamma=4, L = 200$. The dots represent the average of each $(k,\omega)$ point over the 100 runs, while the dashed lines indicate the standard error on the mean. (b) $\widetilde{C}(k,\omega)$ obtained by performing a spline interpolation of $\widetilde{C}_{\rm int}(k,\omega)$ ($k$ fixed) and computing its $\omega$ derivative (bright lines). The noise in the curves is the result of statistical fluctuations from the MCMC sampling, and can be removed by using a simple low-pass filter as described in the text (dark lines).}
    \label{fig:reconstructing_fourier}
\end{figure}

Similarly to \cref{eq:Cxt_estimate} from the main text, $\widetilde{C}_{\rm int}(k,\omega)$ can be estimated on the $k$ and $\omega$ grids via MCMC
\begin{equation}
\label{eq:CkomegaInt_estimate}
    \widetilde{C}_{\rm int}(k,\omega) \ \rightarrow  \ \frac{Z_\lambda}{\ell_{\rm max}}\sum_{\ell=1}^{\ell_{\rm max}}\widetilde{g}^{\rm (int)}_{\blambda,\bmu_\ell}(k,\omega) \qquad \qquad \widetilde{g}^{\rm (int)}_{\blambda,\bmu}(k,\omega) = 2 \pi L \,  \theta_H(\omega -E_\bmu + E_\blambda) \, \delta_{k,P_\bmu-P_\blambda}.
\end{equation}
In \cref{fig:reconstructing_fourier} we plot $\widetilde{C}_{\rm int}(k,\omega)$ obtained in this way for the example of $\tau=5, \gamma=4$ and $L=200$. The data was obtained by averaging the results for each pair $(k,\omega)$ over 100 Markov chains run in parallel, to further reduce statistical fluctuations. At each $k$ value, the curve as a function of $\omega$ appears to be smooth, and the statistical fluctuations around the average are negligible. To obtain $\widetilde{C}(k,\omega) = d\widetilde{C}_{\rm int}(k,\omega)/d\omega$ we compute a numerical derivative by first performing a spline interpolation of $\widetilde{C}_{\rm int}(k,\omega)$ at each $k$, and then taking the derivative of the smooth spline function $f_k(\omega)$. In \cref{fig:reconstructing_fourier}(b) we plot $d f_k(\omega)/d \omega$ (bright lines) obtained in this way from the curves in (a). On top of the smooth profile it is evident the presence of some noise arising from the small but non-zero statistical fluctuations in the values of $\widetilde{C}_{\rm int}(k,\omega)$ (this is a consequence of the fact that, by construction, spline interpolations are designed to pass directly through each given data point). To obtain a smooth Fourier correlator $\widetilde{C}(k,\omega)$ (appropriate for comparisons with $L\to\infty$ limit) we filter out the high-“frequency” components characteristic of the noise (where here “frequency” refers to the Fourier conjugate of $\omega$, i.e. time). This can be achieved by convolution with a low-pass filter that has a low cutoff time $t^*$
\begin{equation}
\begin{aligned}
\label{eq:smoothing}
p_k(t) &= \int_{-\infty}^\infty d \omega \, \frac{d f_k(\omega)}{d \omega} e^{-i \omega t} \ , \\
    \widetilde{C}(k,\omega) &\rightarrow \frac{1}{2\pi}\int_{-\infty}^\infty d t\, p_k(t)\, \theta_H(t^* - |t|) e^{i t \omega} \ .
\end{aligned}
\end{equation}
In practice, we implement the filtering step using the Discrete Fourier Transform (DFT), for computational efficiency. As evident from \cref{fig:reconstructing_fourier}(b) the filtered curves (dark lines) perfectly capture the smooth profile of the original derivatives $d f_k(\omega)/d \omega$. 

We have applied this smoothing procedure to produce all \emph{line} plots of $\widetilde{C}(k,\omega)$ present in the main text and in the appendices. For the \emph{density} plots (both in main text and appendices) we have in most cases instead plotted the output from the numerical derivative of the MCMC data for $\widetilde{C}_{\rm int}(k,\omega)$. The reason for this is that density plots automatically introduce an average (effective smoothing), and are therefore excellent for capturing the main signal in the Fourier correlator already from the raw (non-smoothed) data.

\section{Form factors}
\label{appendix:formfactors}

Consider the form factors $F_{\blambda,\bmu}$ of the Bose field $\phi(x)$ 
\begin{equation}
    F_{\blambda,\bmu} = \braket{\bmu|\phi(0)|\blambda} \ ,
\end{equation}
where $\ket \bmu$ and $\ket \blambda$ are eigenstates normalized to 1 of the LL Hamiltonian. Explicit expression for them can be obtained from the algebraic Bethe ansatz \cite{kojima1997determinant,korepin1993quantum_inverse}. 
A particularly convenient expression has been derived in Refs.~\cite{caux2007one_particle,piroli2015exact_formulas}. 
Consider a set of $N$ rapidities $\lambda_j$ and a set of $N-1$ rapidities $\mu_k$, which satisfy the Bethe equations and such that
\begin{equation}
    \lambda_j \neq \mu_k \qquad \forall \ j , k \ . 
\end{equation}
This condition is generally satisfied due to the nontrivial correlations between rapidities induced by the Bethe equations. Denoting $\lambda_j - \lambda_k$ simply as $\lambda_{jk}$ (and similarly for rapidities $\mu$), the logarithm of $|F_{\blambda,\bmu}|^2$ is given by
\begin{equation}
\label{eq:formfactorBose}
    \begin{aligned}
\ln (|F_{\blambda,\bmu}|^2) = -M_{\blambda,\bmu} &= - \ln\left[\text{det}_N A_\blambda\right]- \ln\left[\text{det}_{N-1} A_\bmu\right]+\sum_{1\le j<k\le N}\ln\left[\lambda_{jk}^2(\lambda_{jk}^2+c^2)\right] \\
& + \sum_{1\le j<k\le N-1}\ln\left[\frac{\mu_{jk}^2}{\mu_{jk}^2+c^2}\right] - \sum_{j=1}^N \sum_{k=1}^{N-1}\ln\left[ (\lambda_j - \mu_k)^2 \right]\\
&+ \sum_{j=1}^N \ln\Big[ \big(2 \operatorname{Im}(V_{\lambda_j})\big)^2 \Big]-\ln\Big[ \big(2 \operatorname{Im}(V_{\lambda_p})\big)^2 \Big]-\ln\Big[ \big(2 \operatorname{Im}(V_{\lambda_s})\big)^2 \Big]\\
& \qquad \qquad \qquad +\ln\Big[\big| \text{det}_{N}(\mathbb{I}+B_{\lambda_s, \lambda_p})\big|^2\Big] \ .
    \end{aligned} 
\end{equation}
Here $\lambda_p$ and $\lambda_s$ are two arbitrary complex numbers, $\mathbb{I}$ is the $N\times N$ identity matrix and
\begin{equation}
   (A_\blambda)_{jk} = \delta_{j,k}\left[ L + \sum_{\ell=1}^N K(\lambda_{j\ell}) \right] - K(\lambda_{jk}) \qquad \qquad j,k = 1, \ldots, N \ ,
\end{equation}
\begin{equation}
    V_{\lambda}=\frac{\prod_{m=1}^{N-1}(\mu_m-\lambda+ic)}{\prod_{m=1}^{N}(\lambda_m-\lambda+ic)} \ ,
\end{equation}
\begin{equation}
\begin{aligned}
    (B_{\lambda_s,\lambda_p})_{jk} &= \frac{1}{2 \operatorname{Im}(V_{\lambda_j})}\frac{\prod_{m=1}^{N-1}\ (\mu_m-\lambda_j)}{\prod_{\substack{m=1 \\ (m \neq j)}}^{N}(\lambda_m-\lambda_j)}\\
   & \ \ \  \times \big[K(\lambda_{jk})-K(\lambda_p - \lambda_j)K(\lambda_s - \lambda_k)\big] \quad \qquad j, k = 1, \ldots, N \ .
\end{aligned}
\end{equation}

The function $K(x)$ is given by $K(x)=2c/(x^2+c^2)$. The matrix $A_\bmu$ is defined identically to $A_\blambda$ but substituting $\blambda \to \bmu$ and restricting the $j,k$ indices from $1$ to $N-1$. Crucially, the final result for $|F_{\blambda,\bmu}|^2$ does not depend on the choice of $\lambda_p$ and $\lambda_s$, which can be chosen arbitrarily \cite{piroli2015exact_formulas}. The formula \eqref{eq:formfactorBose} proves that, fixed any $N$-particle eigenstate $\ket \blambda$, the off-diagonal matrix elements $F_{\blambda,\bmu}=\braket{\bmu |\phi(x)|\blambda}$ of the Bose field $\phi(x)$ are non-zero for \emph{any} $(N-1)$-particle eigenstate $\ket \bmu$ (due to the particle-number selection rule, matrix elements are trivially zero if $\ket \bmu$ has more or less than $N-1$ particles). \\

The expression for $F_{\blambda,\bmu}$ gets drastically simplified \cite{essler2024statistics} in the impenetrable model $c=\infty$. By performing the limit $c\to \infty$ of \eqref{eq:formfactorBose} the divergent terms cancel out and one remains with (we note that the limit $c\to\infty$ is most easily calculated starting from the slightly different expression for $F_{\blambda,\bmu}$ from \cite{caux2007one_particle}, instead of the one from \cite{piroli2015exact_formulas} used in \eqref{eq:formfactorBose}) 
\begin{equation}
\label{eq:formfactoBoseImpenetrable}
\begin{aligned}
    &\ln(|F_{\blambda,\bmu}|^2)\bigg|_{c=\infty} = \sum_{1\le j<k\le N}\ln\left[\lambda_{jk}^2\right]+ \sum_{1\le j<k\le N-1}\ln\left[\mu_{jk}^2\right]
    \\ & \quad \quad- \sum_{j=1}^N \sum_{k=1}^{N-1}\ln\left[ (\lambda_j - \mu_k)^2 \right]  +2 (N-1)\ln 2 - (2N-1)\ln L \ .
    \end{aligned}
\end{equation}
This proves that even if  $c=\infty$ represents a free (fermionic) limit, the matrix elements of the local field $\phi(x)$ between eigenstates of the Hamiltonian (that respect the particle-number selection rule) are not sparse. The reason for this anomalous feature, compared to standard free theories, is that $\phi(x)$ is nonlocal when expressed in terms of the free fermionic field that diagonalizes the $c=\infty$ Hamiltonian \cite{creamer1980quantum}.

We notice that from a computational point of view, numerically evaluating \eqref{eq:formfactoBoseImpenetrable} is significantly faster than evaluating \eqref{eq:formfactorBose}, due to the presence of determinants in the latter.

\section{Exact results in the impenetrable limit from Fredholm determinants}
\label{appendix:fredholm}

In the impenetrable limit $c=\infty$, due to the simpler structure \eqref{eq:formfactoBoseImpenetrable} of the squared form factor $|F_{\blambda,\bmu}|^2$, it is possible to recast the sum over $\ket \bmu$ states that appears in \cref{eq:Cxtdef} for the thermal correlation functions of the Bose field $\phi(x)$ in terms of a Fredholm determinant (see, e.g., \cite{bornemann2010numerical}). This was first achieved for the static case $C(x,t=0)$ by Lenard in \cite{lenard1964momentum,lenard1966one_dimensional}, and then extended to the full $(x,t)$ plane in \cite{korepin1990time_dependent} (see also \cite{kojima1997determinant,patu2008one_dimensional,patu2022exact,korepin1993quantum_inverse}). These closed-form expressions for the correlation functions hold for $L\to\infty$. 

We report here the formulas for the full thermal function $C(x,t)$ and explain how they can be evaluated numerically. We do not report explicitly the (simpler) expressions for the static case $C(x,t=0)$, which can in any case be obtained as $\lim_{t \to 0}C(x,t)$. 
We start with the expression for
\begin{equation}
    R(x,t)= \braket{\phi(0,0)\phi^\dag(x,t)}_T \ ,
\end{equation}
where $\braket{.}_T$ is a thermal expectation value for $L\to\infty$. The correlator $R(x,t)$ can be expressed in terms of a Fredholm determinant (see \cite{korepin1993quantum_inverse})
\begin{equation}
\label{eq:fredholmforRxt}
    R(x,t)=e^{-iht}\left( \frac{1}{2\pi}G(x,t)+\partial_\alpha \right)\det \left(\hat{\mathbb{I}}+\hat{V}(\alpha)\right)\Bigg|_{\alpha = 0}  \ , 
\end{equation}
where $h$ is the chemical potential in \eqref{eq:HsecondQ} and 
\begin{equation}
    G(x,t) = \int_{-\infty}^\infty ds \, \exp(i t s^2 - i x s)  \ .
\end{equation}
$\hat{V}(\alpha)$ is an integral operator acting on a function $f(\lambda)$ as
\begin{equation}
    (\hat{V}(\alpha)f)(\lambda)=\int_{-\infty}^\infty d\mu \, V(\lambda,\mu)f(\mu) \ ,
\end{equation}
with kernel $V(\lambda,\mu)$ defined as
\begin{equation}
    V(\lambda,\mu) = \sqrt{\mathcal{\vartheta(\lambda)}}\,V_0(\lambda,\mu)\sqrt{\mathcal{\vartheta(\mu)}} \qquad \quad \vartheta(\lambda) = \frac{1}{1+e^{(\lambda^2-h)/T}} \ ,
\end{equation}
\begin{equation}
\label{eq:expressionforV0}
\begin{aligned}
    V_0(\lambda,\mu) &= e^{-i t(\lambda^2+\mu^2)/2\, +\, ix(\lambda+\mu)/2}\\
    & \qquad \times \left[ \frac{E(\lambda|x,t)-E(\mu|x,t)}{\pi^2 (\lambda -\mu)} -\frac{\alpha}{2\pi^3}E(\lambda|x,t)E(\mu|x,t)\right]  \ ,
\end{aligned}
\end{equation}
\begin{equation}
\label{eq:Elambdaxt}
    E(\lambda|x,t) = \text{P.V.} \int_{-\infty}^\infty ds\, \frac{\exp(its^2-i xs)}{s-\lambda} \ .
\end{equation}
In \eqref{eq:fredholmforRxt} “$\det$” denotes a Fredholm determinant, $\hat{\mathbb{I}}$ is the operator that acts as the identity $(\hat{\mathbb{I}}f)(\lambda)=f(\lambda)$ and “$\text{P.V.}$” in \eqref{eq:Elambdaxt} indicates the Cauchy principal value. 
To numerically evaluate the Fredholm determinant in \eqref{eq:fredholmforRxt} we follow the standard method from \cite{bornemann2010numerical} based on quadrature rules. In particular, we fix a cutoff $k_{\rm max}$ for the rapidities and use the Gauss-Legendre quadrature rule to create a grid of $M$ points $\lambda_j$, and associated weights $w_j$, in the interval $[-k_{\rm max},k_{\rm max}]$, from which we can approximate the action of the integral operator $\hat{V}(\alpha)$ as
\begin{equation}
    (\hat{V}(\alpha)f)(\lambda_i) \approx \int_{-k_{\rm max}}^{k_{\rm max}} d\mu \, V(\lambda_i,\mu)f(\mu) \approx \sum_{j=1}^M V(\lambda_i,\mu_j)f(\mu_j)\, w_j \ .
\end{equation}
This allows us to interpret, in the limit sufficiently large $M$ and $k_{\rm max}$, the operator $\hat{V}(\alpha)$ as a matrix $\tilde{V}(\alpha)$ with entries
\begin{equation}
    (\tilde{V}(\alpha))_{ij} = V(\lambda_i,\mu_j)\, w_j \ .
\end{equation}
For large $M$ the matrix determinant
\begin{equation}
    \text{det}_M(\mathbb{I} +\tilde{V}(\alpha)) 
\end{equation}
yields an accurate approximation of the Fredholm determinant appearing in \eqref{eq:fredholmforRxt} ($\mathbb{I}$ indicates here the $M\times M$ identity matrix). For the $\partial_\alpha$ derivative in \eqref{eq:fredholmforRxt} we use the following well-known identity for the derivative of the determinant of a matrix $A(\alpha)$
\begin{equation}
    \frac{d}{d\alpha} \det(A(\alpha)) = \det(A(\alpha)) \, \text{Tr}\left[ A^{-1}(\alpha) \frac{d A(\alpha)}{d \alpha} \right] \ ,
\end{equation}
where $\text{Tr}$ denotes the matrix trace. \\

\begin{figure}[t]
    \centering
    \includegraphics[scale=0.34]{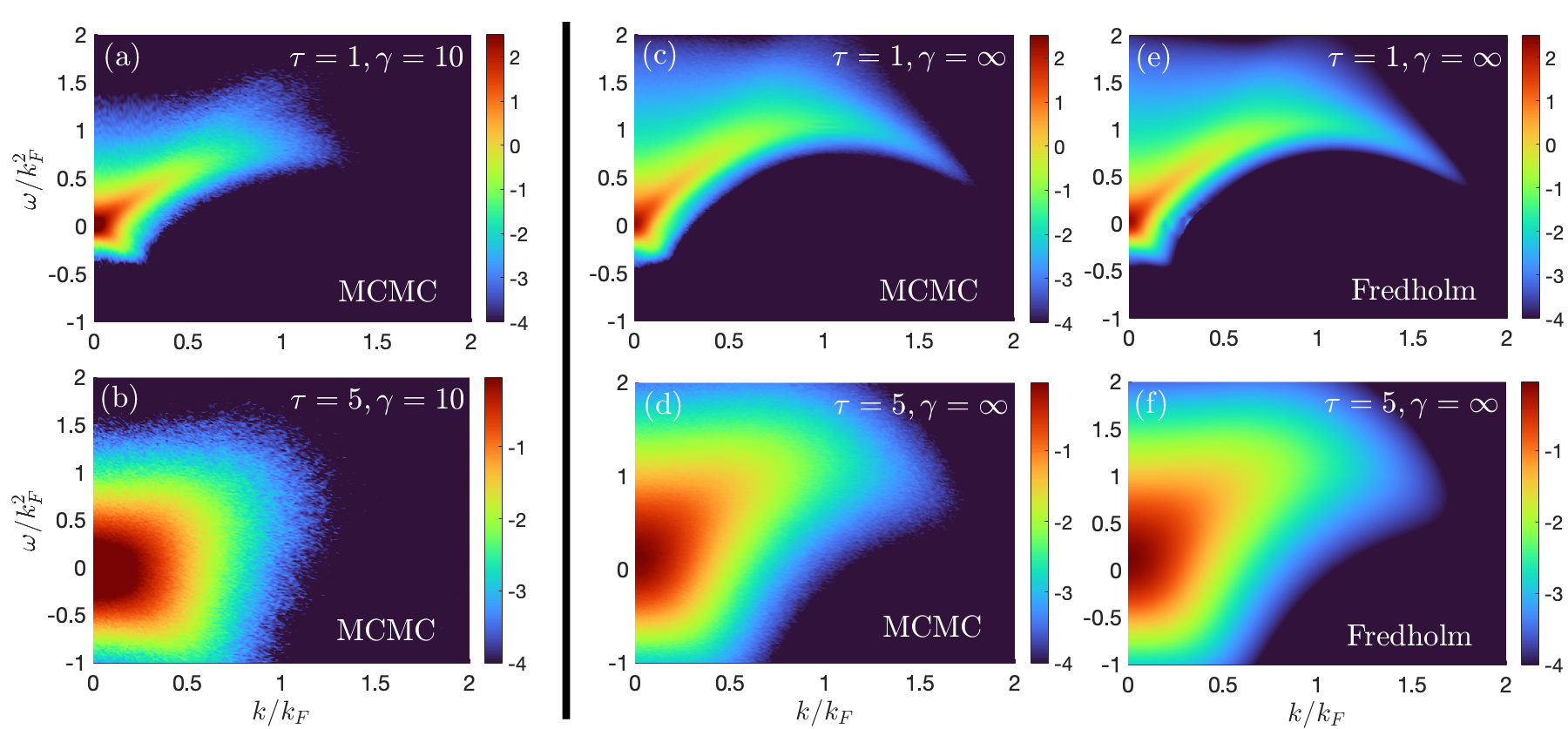}
    \caption{Density plots for $\ln[ \widetilde{C}(k,\omega)/n]$ at $L=200$ ($n=1$) and temperatures $\tau=1$ and $5$. MCMC data obtained by running 100 parallel Markov chains with $\ell_{\rm max}=10^6$--$10^7$ steps each. (a)-(b) MCMC results for large but finite couplings $\gamma=10$. (c)-(d) MCMC results for the impenetrable limit $\gamma=\infty$. (e)-(f) Exact results for $\gamma=\infty$ and $L\to \infty$ from Fredholm representation. The weak “noisy” feature visible in (e) around $k/k_F\approx0.3$ and $\omega \approx 0$ is due to numerical inaccuracies generated by computing Fredholm determinants for the real space correlators $C(x,t)$ and Fourier transforming to $(k,\omega)$ space.}
    \label{fig:final_log_density_plots}
\end{figure}
To obtain the thermal $C(x,t)$ from the expression for $R(x,t)$ we can use the Kubo-Martin-Schwinger (KMS) relation obtained from the cyclic property of the trace
\begin{equation}
    C(x,t) = \braket{\phi^\dag(x,t)\phi(0,0)}_T = \braket{\phi(0,0)\phi^\dag(x,t+i\beta)}_T = R(x,t+i\beta) \ ,
\end{equation}
where $\beta=1/T$. Given that we need to numerically evaluate $M^2$ matrix elements $(\tilde{V}(0))_{ij}$ (for increasingly large $M$, to check convergence), it is convenient to work out an explicit expression for $E(\lambda|x,t + i\beta)$ in \eqref{eq:Elambdaxt}.  We recall the result for the Hilbert transform $\mathcal{H}(f)$ of a Gaussian \cite{king2009hilbert}
\begin{equation}
\label{eq:HilbertransformGaussian}
    f(z) = e^{-a z^2} \qquad \quad \mathcal{H}(f)(z) = -\frac{1}{\pi} \text{P.V.}\int_{-\infty}^\infty ds \, \frac{f(s)}{s-z} = -i e^{-a z^2} \operatorname{erf}(i \sqrt{a}\, z)\ ,
\end{equation}
where $a \in \mathbb{C}$, $\operatorname{Re}(a)>0$, $\sqrt{a}$ has brunch cut along the negative real axis and $\operatorname{erf}$ denotes the error function. The function $E(\lambda|x,t+i\beta)$ can be cast in a very similar form
\begin{equation}
    E(\lambda|x,t+i\beta) = e^{-x^2/(4 \tilde{a})} \ \text{P.V.} \int_{-\infty}^\infty ds\, \frac{\exp\left[-\tilde{a}\left(s+\frac{ix}{2 \tilde{a}}\right)^2\right]}{s-\lambda} \ ,
\end{equation}
with $\tilde{a}=\beta - i t$. Using the simple generalization of \eqref{eq:HilbertransformGaussian} which accounts for shifts like $ix/(2 \tilde{\alpha})$, one obtains the closed-form result
\begin{equation}
    E(\lambda|x,t+i\beta) = i \pi\,e^{-\tilde{a}\tilde{\lambda}^2-x^2/(4 \tilde{a})} \operatorname{erf}(i \sqrt{\tilde{a}}\,\tilde{\lambda}) \qquad \qquad \tilde{\lambda}=\lambda + \frac{ix}{2 \tilde{a}} \ .
\end{equation}
From this we can also obtain the analytic expression for $\partial E(\lambda|x,t+i\beta)/\partial \lambda$, which is needed in \eqref{eq:expressionforV0} to address the limit $|\lambda-\mu| \to 0$.

\section{Additional benchmarks and results}
\label{appendix:additionalbenchandres}
In \cref{fig:final_log_density_plots}(c)-(f) we provide additional benchmarks at temperatures $\tau=1$ and $\tau=5$ for the impenetrable case ($\gamma=\infty$), by showing log-scale density plots for $\widetilde{C}(k,\omega)$ from both MCMC ($L=200$) and exact Fredholm representations ($L\to\infty$). Again, we find excellent agreement. 
In \cref{fig:final_log_density_plots}(a)-(b) we provide similar MCMC results but at finite couplings $\gamma=10$. In \cref{fig:log_density_plots_lower_values} we plot again the MCMC data from \cref{fig:final_log_density_plots}(c) and (d) but down to values of $\ln[\widetilde{C}(k,\omega)]$ of $-8$. We see that the MCMC data appear to capture the signal well even down to these low values, \emph{cf}.~Refs.~\cite{caux2007one_particle, cheng2025one_body} for a direct comparison with log-scale results at $\tau=0$ or very low temperatures. In particular, the $\tau=1, \gamma=\infty$ plots in \cref{fig:final_log_density_plots,fig:log_density_plots_lower_values} feature a spectral weight very similar to the $\tau=0$ results of \cite{caux2007one_particle}, with a type II dispersion \cite{lieb1963exact_2} visible at $\omega>0$. 
\begin{figure}[t]
    \centering
    \includegraphics[scale=0.3]{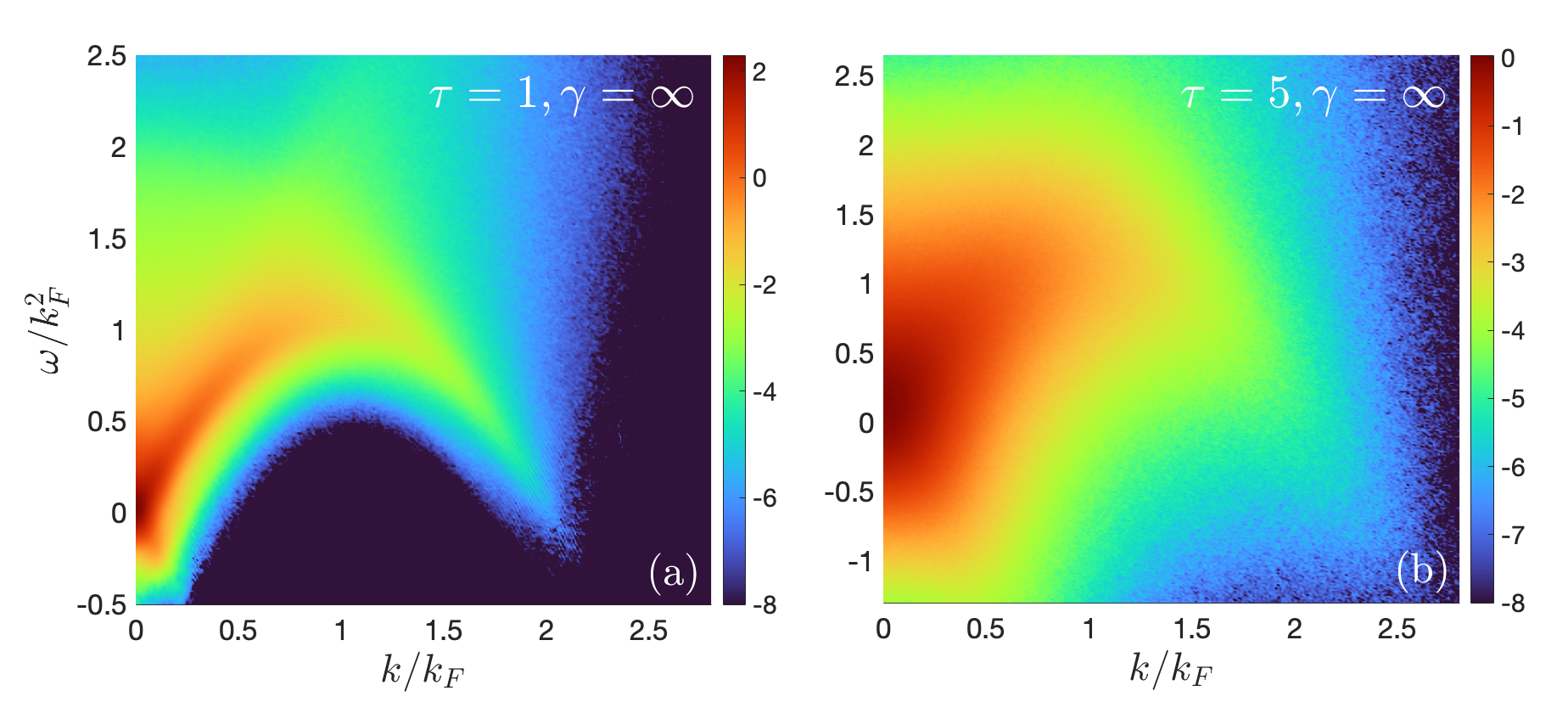}
    \caption{Density plots for $\ln[ \widetilde{C}(k,\omega)/n]$ for the same MCMC data as \cref{fig:final_log_density_plots}(c) and (d) but on a colour scale down to $-8$.}
    \label{fig:log_density_plots_lower_values}
\end{figure}
This zero-temperature dispersion $\omega_{\rm 2}(k)$ is associated with energies $\omega_2=E_\bmu-E_\blambda$ and momenta $k=P_\bmu -P_\blambda$ where $\ket \blambda$ is the ground state of $N$ particles and $\ket \bmu$ is constructed by starting from the ground state with $N-1$ particles and implementing a particle-hole excitation that creates a hole in the Fermi sea and adds a particle at the positive boundary of the latter, i.e.~just to the right of the Fermi momentum. For any interaction coupling $c>0$ the dispersion $\omega_2(k)$ vanishes at $k=0$ and $k=2k_F$ \cite{lieb1963exact_2} and is maximal at $k=k_F$, where $k_F=n \pi$ denotes the Fermi momentum of the impenetrable $c = \infty$ model. Given that at $\tau=0$ the type II dispersion $\omega_2(k)$ represents a \emph{sharp lower threshold} for $\widetilde{C}(k,\omega)$ \cite{caux2007one_particle} (\emph{cf.} also \cref{fig:T_0}), at small but non-zero $\tau$ this feature is expected to remain visible, although it can be broadened by the finite temperature. This is verified in \cref{fig:final_log_density_plots}(a) and (c), and in \cref{fig:log_density_plots_lower_values}(a), which  shows that the signal stretches up to $2 k_F$. 
\begin{figure}
    \centering
    \includegraphics[width=0.4\linewidth]{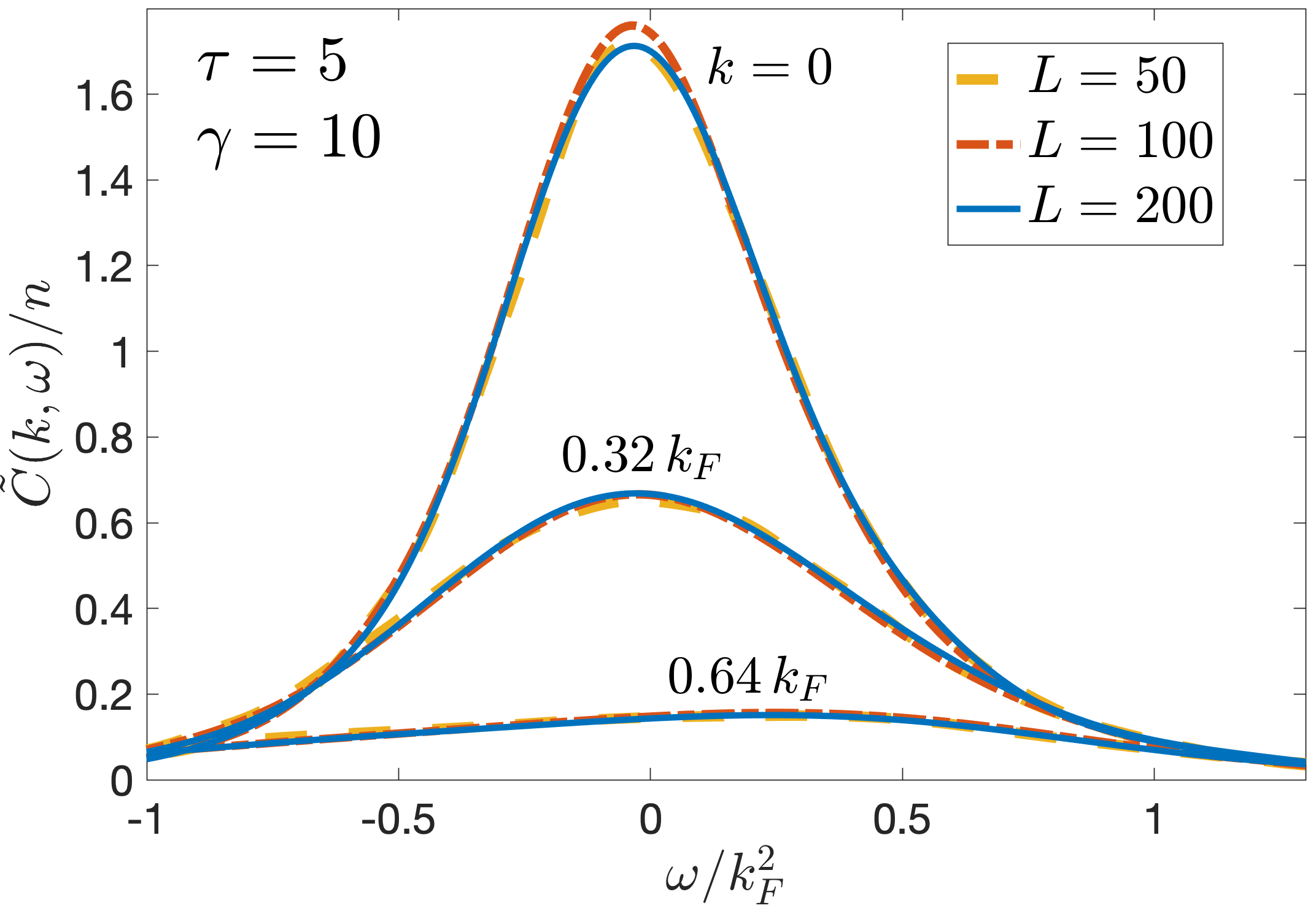}
    \caption{Finite size effects in the values of $\widetilde{C}(k,\omega)$, at $\tau=5$, $\gamma=10$ ($n=1$). The frequency and momenta are rescaled with $k_F=\pi \, n$.}
    \label{fig:finite_size_effect_SM}
\end{figure}
\begin{figure}[t]
    \centering
    \includegraphics[width=0.95\linewidth]{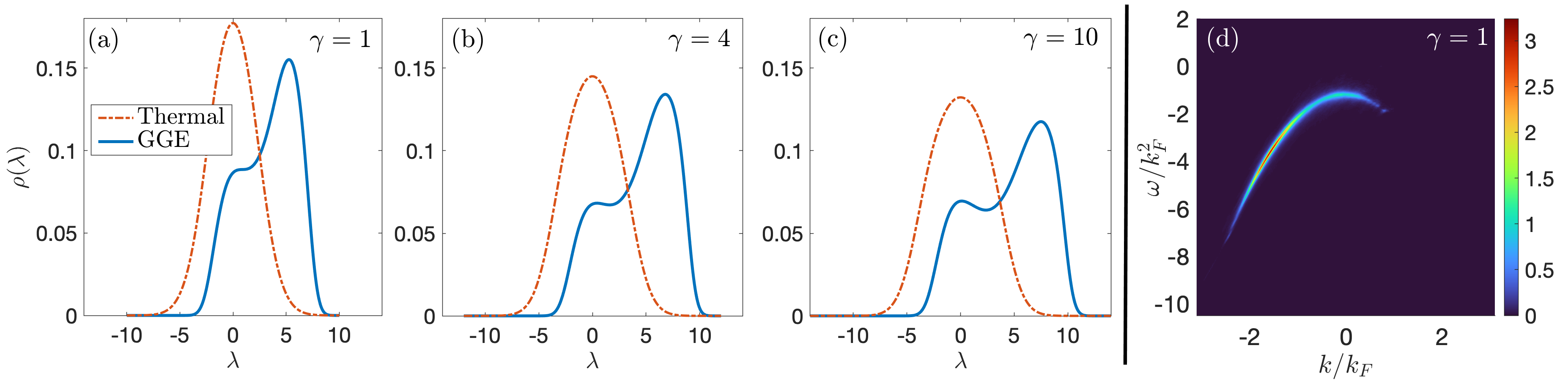}
    \caption{(a)-(c) Root densities $\rho(\lambda)$ for GGE and thermal ensembles used, respectively, in \cref{fig:GGE} and in some of the subplots of \cref{fig:stationary,fig:final_density_plots,fig:finite_size_effect}. The values of the chemical potentials $\beta_j$ for the GGEs in (b) and (c) are those reported in the caption of \cref{fig:GGE}. In (a), for $\gamma=1$, we have set: $T = 1/\beta_2=10$, $h = -\beta_0/\beta_2 =-8.8$ (chosen to set $n=N/L=1$), $\beta_3 = -0.051$, $\beta_4 = 0.0053$ and $\beta_j = 0$ for $j =1$ and $j\ge 5$. (d) Density plot for $\widetilde{C}(k,\omega)/n$ where $\ket \bla$ is associated with the GGE in (a), at $L=200$ (\emph{cf.}~\cref{fig:GGE} for comparison with higher $\gamma$).}
    \label{fig:GGE_SM}
\end{figure}
In \cref{fig:finite_size_effect_SM} we report additional data for estimating finite size effects on the values of $\widetilde{C}(k,\omega)$, here at $\tau=5$, $\gamma=10$. Similarly to \cref{fig:finite_size_effect} from the main text, in the range of $k$ and $\omega$ values of interest finite size effects are very small already at $L\gtrsim 50$. 

In \cref{fig:GGE_SM}(a)-(c) we compare the root densities $\rho(\lambda)$ for some of the GGE and thermal macrostates discussed in the main text, in particular those for which $T=1/\beta_2=10$ in \eqref{GGE}. For the GGEs we choose $\beta_3 \neq 0$, which breaks the symmetry around $0$ (parity) that characterizes the thermal root densities. \cref{fig:GGE_SM}(d) presents a density plot of $\widetilde{C}(k,\omega)$ for $\gamma=1$ and $\ket \bla$ associated with the GGE in \cref{fig:GGE_SM}(a). Exactly as seen for the thermal density plots in the main text (\emph{cf.}~\cref{fig:final_density_plots}), in this low $\gamma$ limit we obtain a sharp signal that is very close to the parabolic dispersion $\omega(k)=-k^2+h$ of the free limit $\gamma=0$ (note that eactly at $\gamma=0$ non-zero values appear only at minus the occupied momenta of the $\ket \bla$ eigenstate, i.e.~at $k_i=-\lambda_i$). The slight broadening is due to the finite temperature and non-zero $\gamma$. By comparison with the higher-$\gamma$ density plots in \cref{fig:GGE} we see how an increase of the interaction strength leads to further broadening of the signal.\\

As a final benchmark we have applied our MCMC scheme to the ground-state dynamical correlator $(T=0)$, which was previously obtained in Ref.~\cite{caux2007one_particle}
by performing an explicit sum over relevant eigenstates $\ket \bmu$ in the Lehmann representation.
The results for $\ln \widetilde{C}(k,\omega)$ at $\gamma=4$, reported in Fig.~\ref{fig:T_0} are in excellent agreement with Fig.~1 of \cite{caux2007one_particle}. In particular, we observe a perfectly sharp lower threshold that coincides with the type II dispersion $\omega_2(k)$, while the central peak in the weight can be thought of as a broadened (and lowered in energy) type I dispersion \cite{lieb1963exact_2} (around which the peak is centered at lower values of $\gamma$ \cite{caux2007one_particle}). We note that the limit $T=0$ necessarily introduces a degree of “lumpiness” (similar to the one discussed in the next section) in the form factors, which translates into a lower MCMC acceptance rate (around $3$-$4\%$) compared to the $T\gg1$ case (in excess of $10\%$). In the current MCMC framework (developed to address finite temperatures), this lower acceptance rate reflects the necessity of the Markov chain to spend significant “time” in the eigenstates associated with the (by far) highest form factors (only in this way the sampling can capture their significantly higher value compared to other form factors). We note that to slightly increase the rate at $T=0$, it is possible to replace the Metropolis algorithm (symmetric proposal matrix $w$) with a Metropolis-Hastings algorithm (non-symmetric $w$) which proposes more often moves that involve at least one of the two boundaries of the Fermi sea present at $T=0$. We remark that the presence of lumpiness at $T=0$ is directly related to the reason why it is in the first place possible to perform an \emph{explicit} Lehmann sum, as in \cite{caux2007one_particle}, for values of $N\lesssim200$ (see next appendix). 
\begin{figure}
    \centering
    \includegraphics[width=0.7\linewidth]{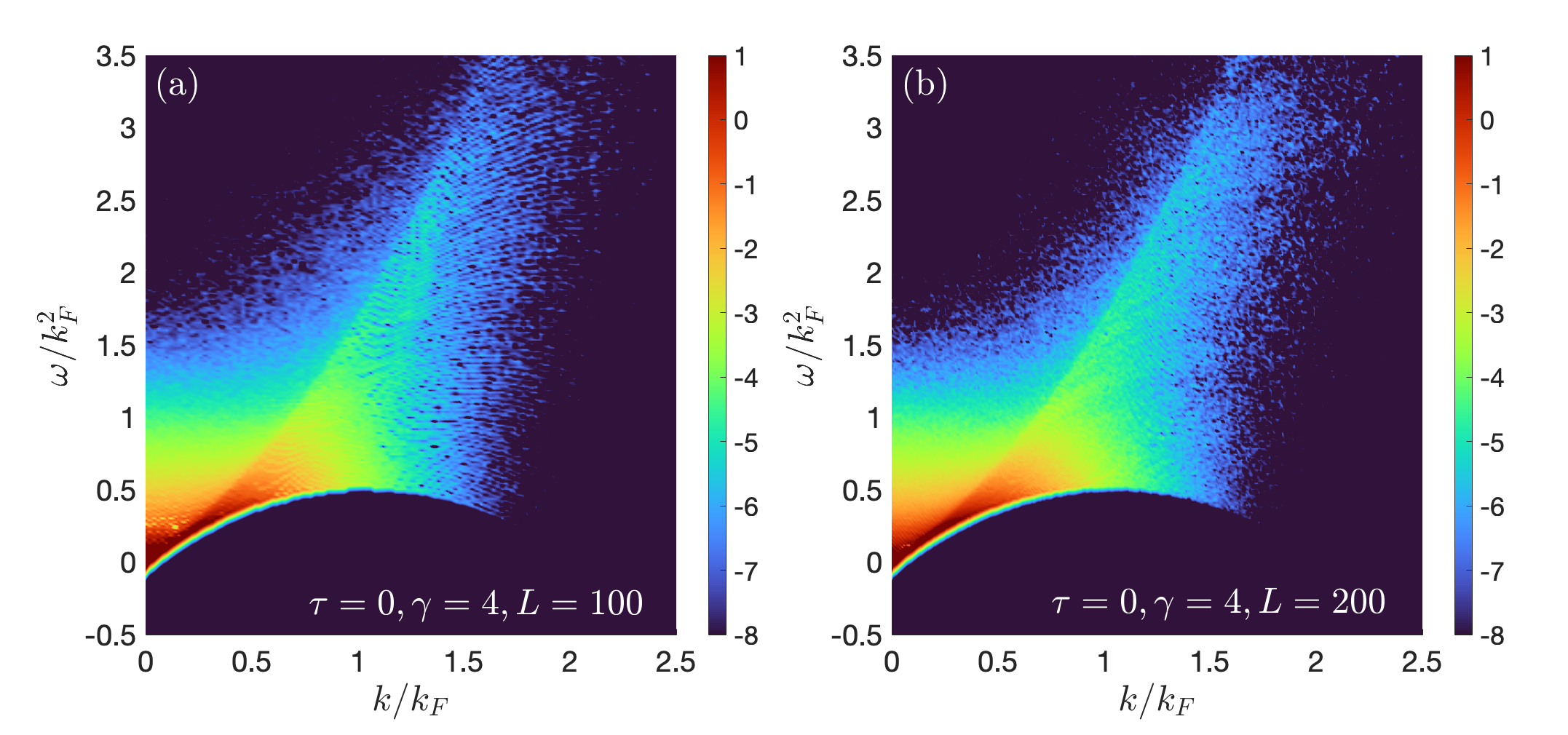}
    \caption{Density plots at $T=0$ ($n=1$) for $\ln[ \widetilde{C}(k,\omega)/n]$. The system sizes are $L=100$ (a), $L=200$ (b). MCMC data obtained by running 200 parallel Markov chains with $\ell_{\rm max}=10^6$ steps each. The “granularity” visible is due to the fact that we are producing the density plot directly from the numerical derivative of $\widetilde{C}_{\rm int}(k,\omega)$ (see \eqref{eq:CkomegaInt_estimate}) obtained from the MCMC data, without implementing the smoothing procedure discussed in \cref{eq:smoothing}. This is because the density plot already introduces an effective smoothing, and the physical signal is perfectly visible already from the raw data.}
    \label{fig:T_0}
\end{figure}

\section{Bose field vs density operator two-point functions}
\label{eq:densityvsBosefield}

Local operators in interacting integrable theories possess exponentially many (in system size $L$) \emph{non-zero} matrix elements between eigenstates \cite{essler2024statistics,slavnov1989calculation,slavnov1990nonequal,kojima1997determinant,smirnov1992form,korepin1993quantum_inverse,caux2007one_particle,piroli2015exact_formulas,brenes2020low,zhang2022statistical,leblond2019entanglement,leblond2020eigenstate,rottoli2025eigenstate}. The vast majority of them is expected to be \emph{at least} exponentially suppressed in $L$ \cite{essler2024statistics, rottoli2025eigenstate} (see below for an exception). On the contrary, in free theories, only a few “sparse” matrix elements generally do not vanish, and their magnitudes decay to zero at most polynomially with $L$.

Exactly as for the Bose field $\phi(x)$, in the interacting regime ($0<c<\infty$) of the LL model the exponential (or super-exponential) suppression of the matrix elements of the density $\varrho(x)=\phi^\dag(x)\phi(x)$ can be easily checked numerically starting from known expression for the density form factors \cite{slavnov1989calculation,slavnov1990nonequal}, or analytically by employing expansions of the form factors in $1/c$ (for $c$ large but finite) \cite{essler2024statistics}. This suppression implies that, as we increase $L$ (at fixed $N/L$), the use of sampling schemes for the reconstruction of correlations functions becomes eventually \emph{necessary} for both $\phi(x)$ and $\varrho(x)$, given that the number of relevant terms becomes prohibitively large at sufficiently large $L$. However, the values of $L$ beyond which such schemes are required can, and do, depend on the operator under consideration.
To show this, we highlight here important differences in the structure of matrix elements of $\phi(x)$ and $\varrho(x)$ at “intermediate” values of $L$, which are perhaps not surprising in light of the fact that matrix elements of $\varrho(x)$ are sparse both at $c=0$ and $c=\infty$, while those of $\phi(x)$ correspond to a standard free theory only at $c=0$ (as already discussed in the main text). These differences shed some light on why correlation functions of $\varrho(x)$ can be reconstructed, even at non-zero temperatures, up to fairly large system sizes ($N\approx50$--$100$) \cite{panfil2014finite} without the use of sampling, although sampling becomes necessary at larger $L$. En passant, we uncover the existence of some matrix elements of $\varrho(x)$ that are merely polynomially suppressed in $L$ (as in free theories) even in the interacting regime $0<c<\infty$ (see also \cite{de_nardis2019diffusion}). This is an interesting fact by itself.
\begin{figure}[!t]
    \centering
    \includegraphics[width=0.9\linewidth]{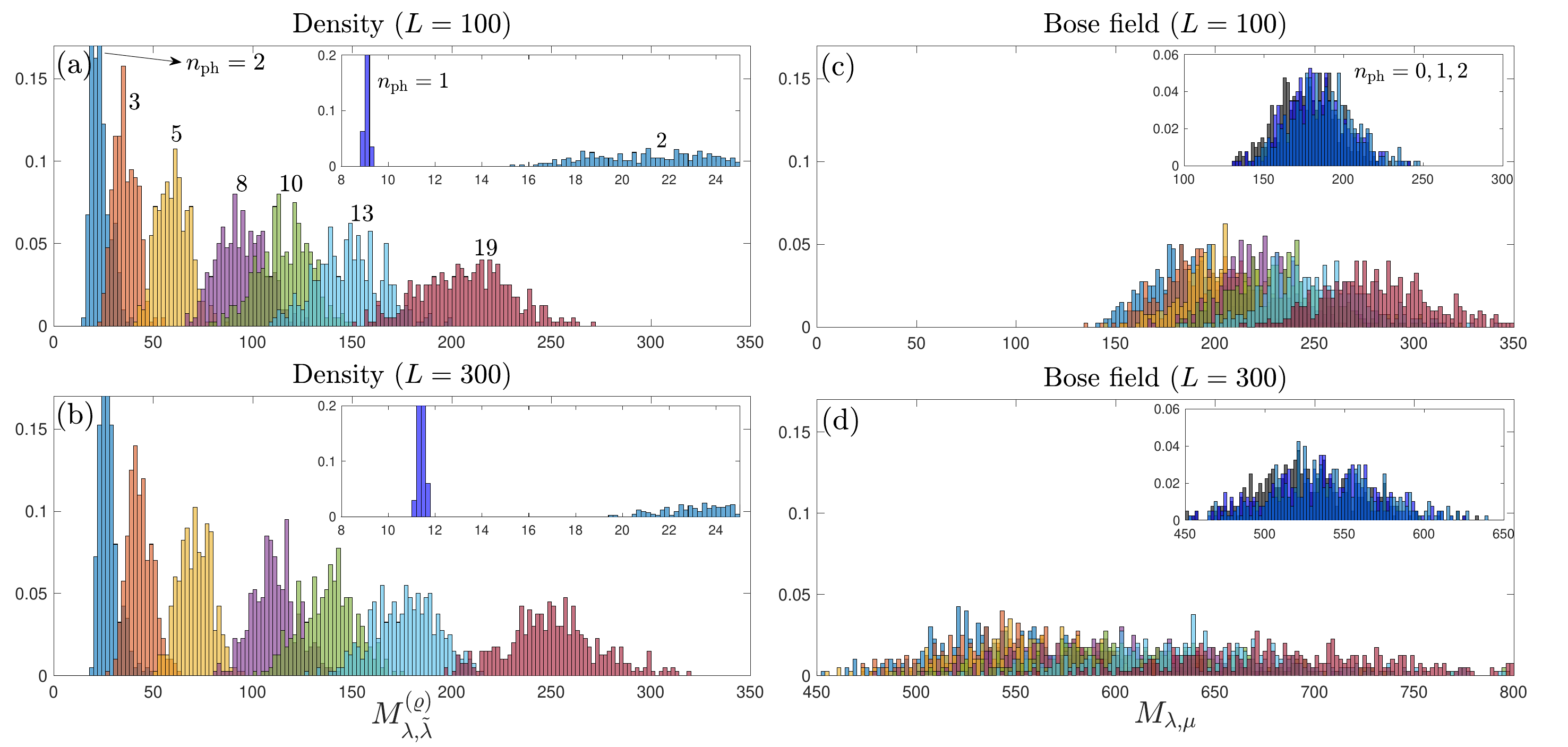}
    \caption{Histograms for $M_{\blambda,\tilde{\blambda}}^{(\varrho)}$ (density) and $M_{\blambda,\bmu}$ (Bose field) from randomly drawn eigenstates $\ket{\tilde{\blambda}}$, $\ket \bmu$ at $\tau = 500$, $\gamma=7$ (for density $n=N/L=1$) and sizes $L=100,300$. Each $\ket{\tilde{\blambda}}$ and $\ket \bmu$ is generated by applying $n_{\rm ph}$ random particle-hole excitations on top of, respectively, $\ket \blambda$ and $\ket{\bmu_{\rm ref}}$ (see text for the construction of $\ket{\bmu_{\rm ref}}$). Histograms display the relative frequency over the 400 eigenstates drawn for each $n_{\rm ph}$. Each particle-hole excitation is constrained to have momentum transfer $|\Delta P_{\rm ph}|\le 1$. Colours from blue to red are associated with $n_{\rm ph}=2,3,5,8,10,13,19$, see (a). Insets show a zoom-in on the region with the lowest number of particle-hole excitations: $n_{\rm ph}=1,2$ for the density in (a)-(b) and $n_{\rm ph}=0,1,2$ for the Bose field in (c)-(d) (the value $n_{\rm ph}=0$ indicates the reference state $\ket{\bmu_{\rm ref}}$).}
    \label{fig:histograms_T_500_c_7}
\end{figure}

Given a reference eigenstate $\ket \blambda$ of $N$ particles, the form factors 
\begin{equation}
\label{eq:formfactorrho}
    |F_{\blambda,\tilde{\blambda}}^{(\varrho)}|^2 = |\braket{\tilde{\blambda}|\varrho(0,0)|\blambda}|^2 = \exp (-M_{\blambda,\tilde{\blambda}}^{(\varrho)})
\end{equation}
of the density are non-zero for any $N$-particle $\ket{\tilde{\blambda}}$ with $P_\blambda \neq P_{\tilde{\blambda}}$ \cite{slavnov1989calculation,slavnov1990nonequal} (we are restricting to off-diagonal matrix elements, i.e.~we require $\ket{\tilde{\blambda}} \neq \ket \blambda$). Similarly to the case of $M_{\blambda,\bmu}$ for $\phi(x)$, $M_{\blambda, \tilde{\blambda}}^{(\varrho)}$ scales super-extensively with $L$ for typical eigenstates $\ket{\tilde{\blambda}}$ in the same macrostate as $\ket \blambda$ and eigenstates in different macrostates \cite{essler2024statistics}. Hence, for large $L$ only rare atypically large form factors in the same macrostate as $\ket \blambda$ contribute a finite amount to correlations functions. Below we focus on these. When evaluating form factors we fix $\ket \blambda$ to be
a typical $N$-particle eigenstate ($N$ even) corresponding to a thermal macrostate at $\tau=T/n^2$ and $\gamma=c/n$. As in the main text, $\ket \blambda$ is chosen to be a “smooth microstate” with respect to the root density $\rho(\lambda)$ of the macrostate chosen, i.e.~we require the finite-$N$ distribution of its rapidities to smoothly match the profile of $\rho(\lambda)$. To meaningfully compare matrix elements of the two different operators we need to draw a parallelism between the $N$-particle eigenstates $\ket{\tilde{\blambda}}$ for the density (characterized by half-odd integer quantum numbers $\{J^{(\tilde{\blambda})}_j\}$) and the $(N-1)$-particle eigenstates $\ket \bmu$ for the Bose field (characterized by integers $\{J^{(\bmu)}_j\}$). We start by considering the limit of large $\tau$, which corresponds to high temperatures $T$ or low densities of particles $n$. Here, each half-odd integer in the set of quantum numbers $\{I_j^{(\blambda)}\}$ that fix $\ket \blambda$ is well separated from its nearest neighbours. This allows us to construct a reference $(N-1)$-particle eigenstate $\ket{\bmu_{\rm ref}}$ by removing an element from $\{I_j^{(\blambda)}\}$ (we always choose this among the two half-odd integers of lowest absolute value) and shifting all the remaining $I_j^{(\blambda)}$ by $\pm 1/2$, where each sign is chosen at random. Note that in the limit of large $\tau$ (and large $\gamma$) it is easy to check that states $\ket{\bmu_{\rm ref}}$ are associated with the lowest possible factors $M_{\blambda,\bmu}$ (i.e.~the largest matrix elements of the Bose field).  We can generate arbitrary eigenstates $\ket{\tilde{\blambda}}$ and $\ket \bmu$ by adding an arbitrary number $n_{\rm ph}$ of particle-hole excitations on top of, respectively, $\ket \blambda$ and $\ket{\bmu_{\rm ref}}$.
\begin{figure}[!t]
    \centering
    \includegraphics[width=0.9\linewidth]{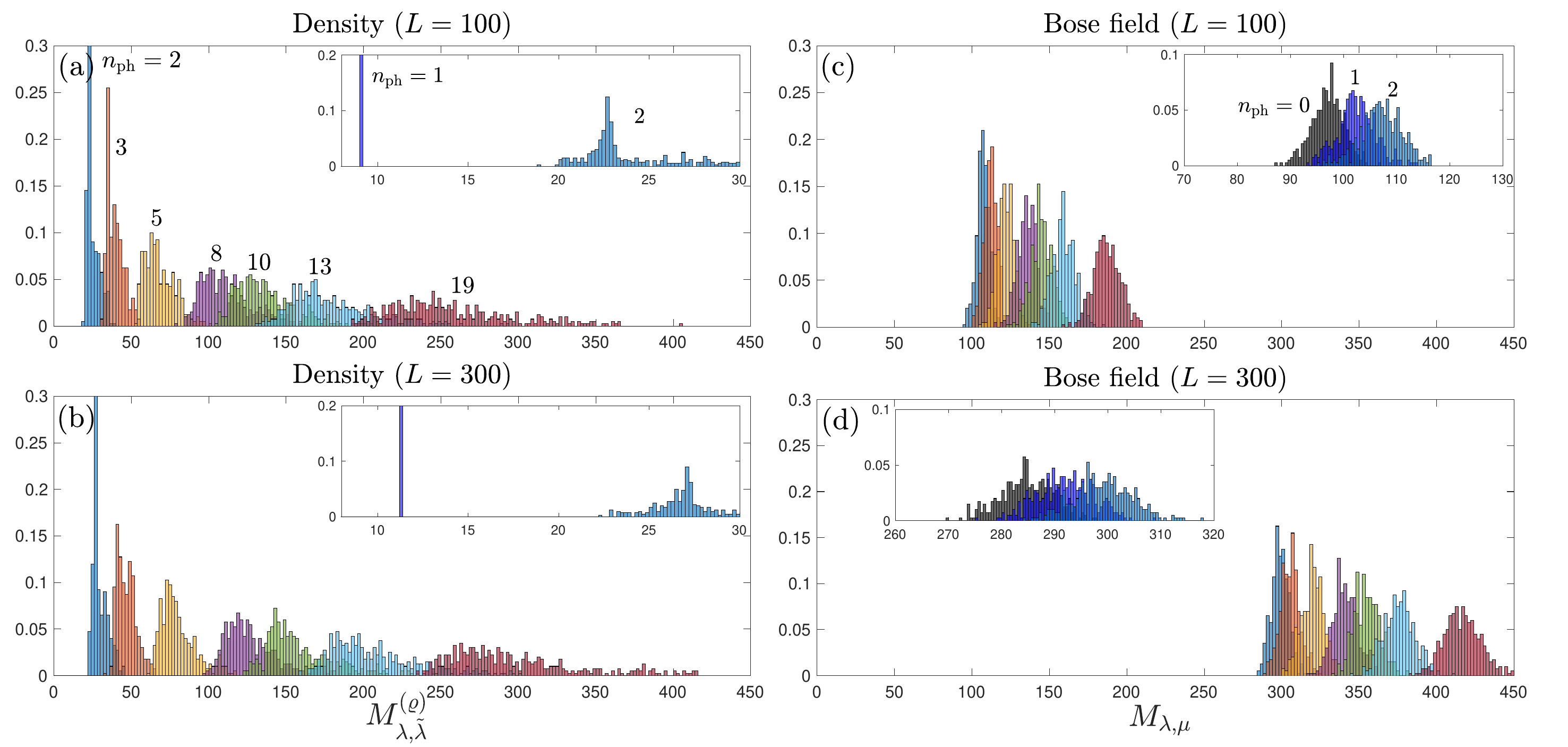}
    \caption{Same as \cref{fig:histograms_T_500_c_7} but for the higher coupling strength of $\gamma=70$ ($\tau=500$).}
    \label{fig:histograms_T_500_c_70}
\end{figure}
\begin{figure}[!b]
    \centering
    \includegraphics[width=0.9\linewidth]{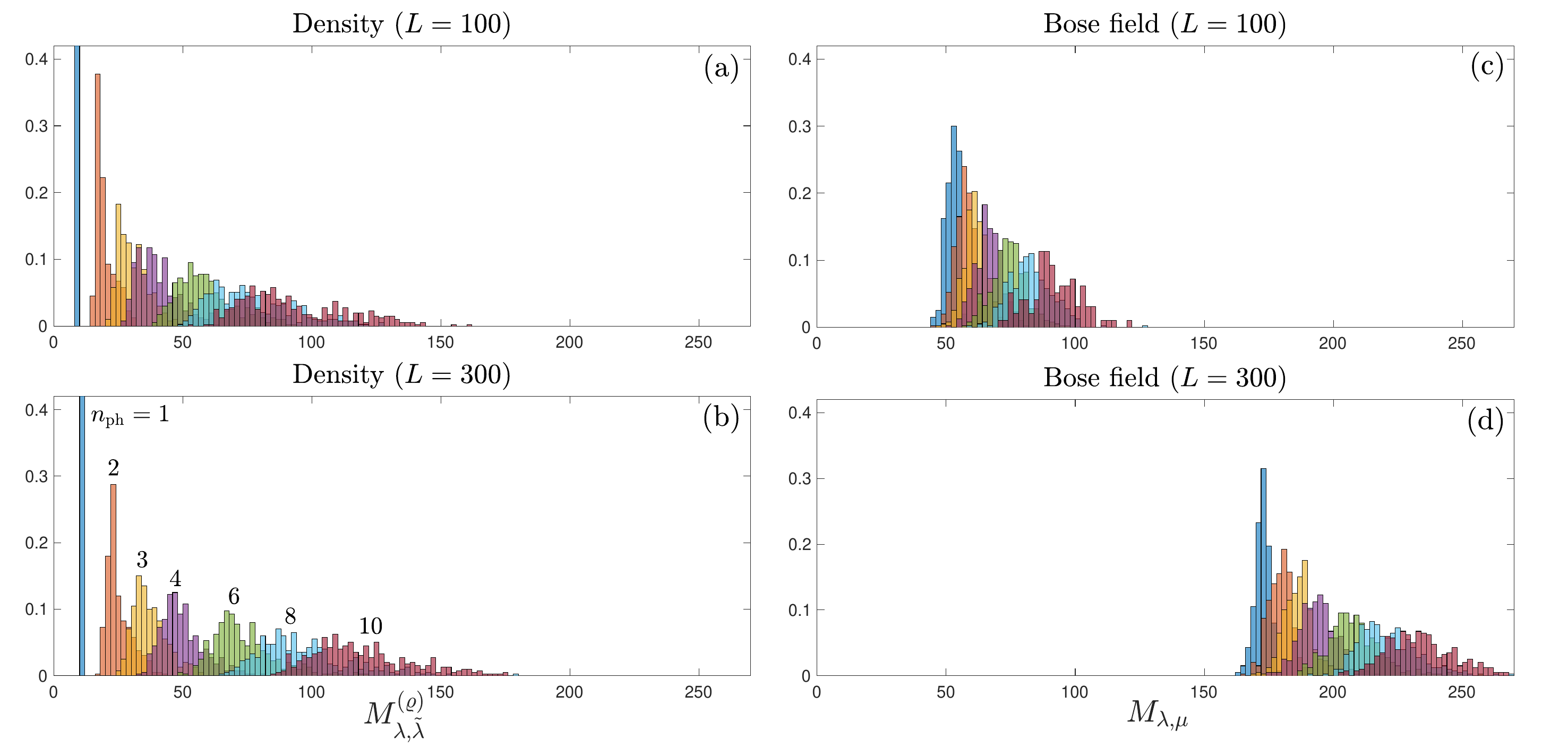}
    \caption{Same as \cref{fig:histograms_T_500_c_7,fig:histograms_T_500_c_70} but for the lower dimensionless temperature $\tau=5$, at coupling strength $\gamma=7$. Colours from blue to red are associated with values $n_{\rm ph}=1,2,3,4,6,8,10$, see (b).}
    \label{fig:histograms_T_5_c_7}
\end{figure}
By a particle-hole excitation we mean the replacement a (half-odd) integer from a set of quantum numbers $\{J\}$ with an another \emph{unoccupied} (half-odd) integer. This construction allows us to compare $M_{\blambda,\bmu}$ and $M_{\blambda,\tilde{\blambda}}^{(\varrho)}$ for eigenstates $\ket{\bmu}$ and $\ket{\tilde{\blambda}}$ arising from an equal number $n_{\rm ph}$ of particle-hole excitations. In \cref{fig:histograms_T_500_c_7} we show histograms for 400 randomly drawn eigenstates $\ket{\tilde{\blambda}}$, $\ket \bmu$ for different values of $n_{\rm ph}$ at $\tau=500$ and $\gamma=7$. We find that the results for the density display a considerable degree of “lumpiness” (i.e.~multimodality of the underlying distribution) compared to the ones for the Bose field. Indeed, for $\varrho(x)$ the peaks associated with different values of $n_{\rm ph}$ are well separated from each other, and the lumpiness appears to become more significant as $L$ is increased. The mode separation is most apparent for low values of $n_{\rm ph}$, see insets in \cref{fig:histograms_T_500_c_7}(a)-(b). On the other hand, it is significantly more difficult to resolve different peaks in the case of the Bose field, and the degree of multimodality decreases as we increase $L$. Even more striking is the difference in the \emph{magnitude} of the $M_{\blambda,\tilde{\blambda}}^{(\varrho)}$ and $M_{\blambda,\bmu}$ sampled. Indeed, $M_{\blambda,\bmu}$ is associated with values significantly larger than $M_{\blambda,\tilde{\blambda}}^{(\varrho)}$, both with respect to a given $n_{\rm ph}$ and overall (in this regard, notice the different $x$-axis scale in \cref{fig:histograms_T_500_c_7}(d)). Furthermore, $M_{\blambda,\tilde{\blambda}}^{(\varrho)}$ appears to vary very slowly with $L$, unlike $M_{\blambda,\bmu}$ which is subject to changes of order $L$.  This dependence on $L$ already suggests that the number of relevant states for the reconstruction of correlation functions of $\phi(x)$ is \emph{significantly higher} (at fixed $L$) than for the case of $\varrho(x)$. In \cref{fig:histograms_T_500_c_70} we present analogous plots for the higher coupling $\gamma=70$ (for the same $\tau=500$ of \cref{fig:histograms_T_500_c_7}). Here the multimodality in the histograms is evident also for the Bose field, but the degree of lumpiness remains considerably higher for the matrix elements of the density (as particularly evident from the insets). Also in this case we observe that values of $M_{\blambda,\bmu}$ are considerably higher than their corresponding $M_{\blambda,\tilde{\blambda}}^{(\varrho)}$, and the matrix elements of the density appear to decrease very mildly with $L$ (unlike those of the Bose field). It is also evident that the sampled values of $M_{\blambda,\bmu}$ cluster within a smaller region than the $M_{\blambda,\tilde{\blambda}}^{(\varrho)}$ ones.

To address a region of the $\tau$-$\gamma$ parameter space closer to the one that, for $\varrho(x)$, has been probed by the ABACUS algorithm \cite{panfil2014finite} we want to perform similar comparisons at lower values of $\tau$ (corresponding to lower temperatures or higher density of particles). In this case, constructing $\ket{\bmu_{\rm ref}}$ by the previous method based on the $\pm 1/2$ perturbations is not as meaningful as in the high-$\tau$ regime. Indeed, at lower $\tau$ most half-odd integers in $\{I_j^{(\blambda)}\}$ lie very close to their nearest neighbours and random choices of the signs $\pm $ typically lead to double occupancies (which must be discarded). Instead of focusing on $\pm 1/2$ perturbations that avoid the previous issue (by more carefully selecting the signs $\pm$), here we choose $\ket{\bmu_{\rm ref}}$ to be the smooth microstate of $(N-1)$ particles associated with the root density $\rho(\lambda)$ that specifies the macrostate of $\ket \blambda$. Note that unlike before, we now have a single $\ket{\bmu_{\rm ref}}$ at each $L$ instead of a family of them. In \cref{fig:histograms_T_5_c_7} we plot the histograms of $M_{\blambda,\tilde{\blambda}}^{(\varrho)}$ and $M_{\blambda,\bmu}$ at $\tau=5$ and $\gamma=7$. We see that the features highlighted in the high-$\tau$ regime, and in particular the marked difference in the magnitudes of $M_{\blambda,\tilde{\blambda}}^{(\varrho)}$ and $M_{\blambda,\bmu}$ (and their scaling with $L$), remain true also here. 

\begin{figure}[!t]
    \centering
    \includegraphics[width=0.9\linewidth]{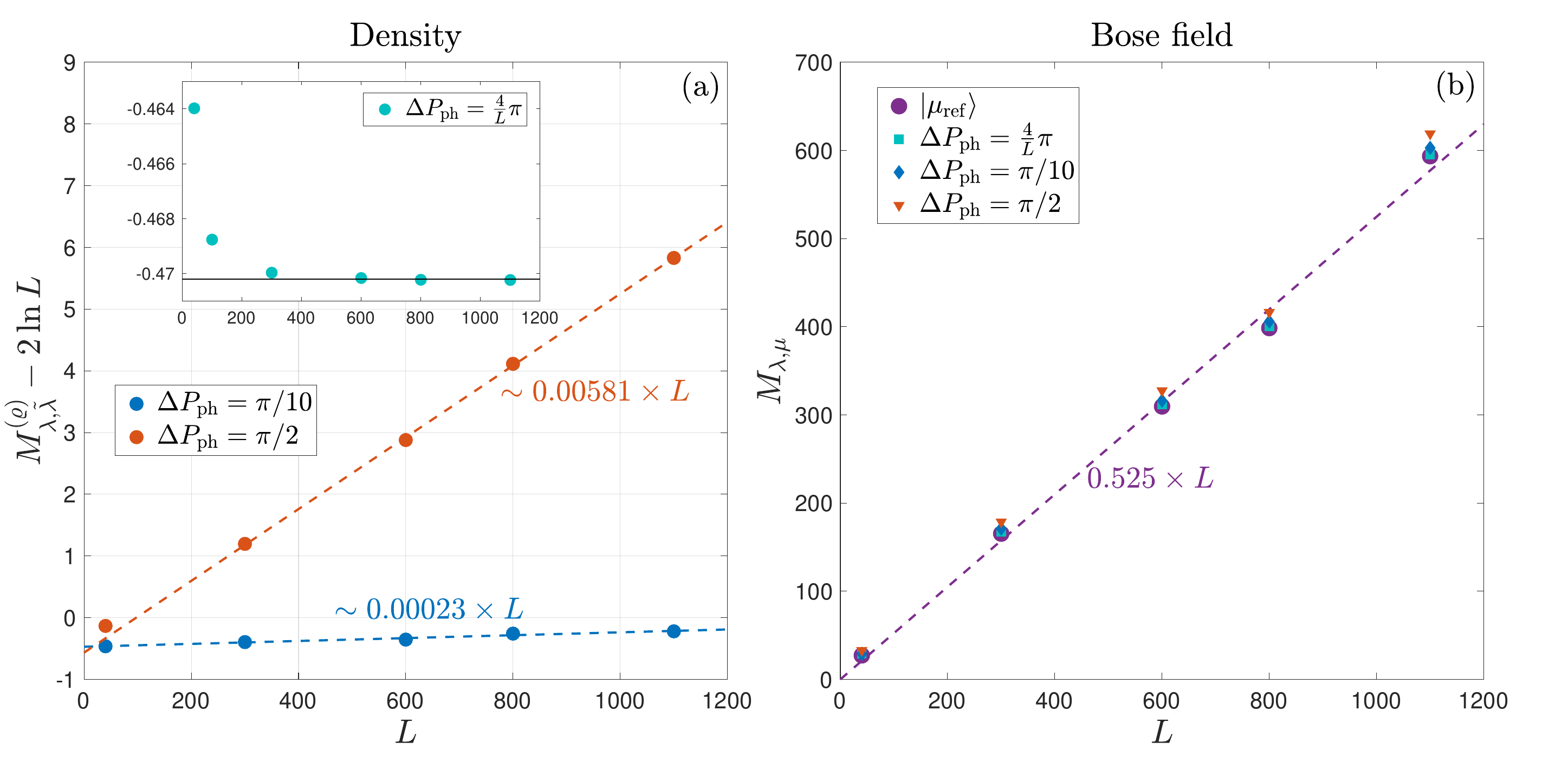}
    \caption{Scaling of $M_{\blambda,\tilde{\blambda}}^{(\varrho)}$ and $M_{\blambda,\bmu}$ with $L$ at $\tau=5, \gamma=7$. The eigenstates $\ket{\tilde{\blambda}}$ and $\ket \bmu$ are constructed by implementing a single particle-hole excitation of momentum $\Delta P_{\rm ph}$ on top of, respectively, $\ket \blambda$ and $\ket{\bmu_{\rm ref}}$. (a) $M_{\blambda,\tilde{\blambda}}^{(\varrho)}-2 \ln L$  as a function of $L$, following the functional form conjectured in \cite{essler2024statistics}. Dashed lines represent 2-parameter linear fits $a\, L + b$ (performed on the 4 data points of largest $L$). The inset shows analogous data for the vanishing momentum $\Delta P_{\rm ph}=4\pi/L$, with the horizontal black line denoting the constant asymptotic value. (b) $M_{\blambda,\bmu}$ as a function of $L$ for the same momenta $\Delta P_{\rm ph}$ of (a) and for the reference $(N-1)$-particle state $\ket{\bmu_{\rm ref}}$ (see text). The dashed line indicate a 1-parameter linear fit $a \, L$ on the $\ket{\bmu_{\rm ref}}$ data (performed on the 4 data points of largest $L$), for which we obtain $a \approx 0.525$. Linear fits for the $\Delta P_{\rm ph}>0$ data (not reported) feature a parameter $a$ that grows with $\Delta P_{\rm ph}$, e.g.~we find $a \approx 0.550$ for $\Delta P_{\rm ph}=\pi/2$.}
    \label{fig:scaling}
\end{figure}

To explore in more detail the different scalings with $L$ of $M_{\blambda,\tilde{\blambda}}^{(\varrho)}$ and $M_{\blambda,\bmu}$ we focus on single particle-hole excitations, i.e.~we fix $n_{\rm ph} = 1$. We consider again $\tau=5$, hence $\ket{\bmu_{\rm ref}}$ is the smooth microstate with $(N-1)$ particles (note that $P_{\bmu_{\rm ref}}=P_{\blambda}=0$). For the density (Bose field) we implement a particle-hole excitation on top of the reference state $\ket \blambda$ ($\ket{\bmu_{\rm ref}}$) by removing one among a few half-odd integers (integers) of lowest absolute value and inserting an unoccupied positive one, such to create a positive momentum transfer $\Delta P_{\rm ph}>0$. 
In \cref{fig:scaling}(a) we present the scaling of $M_{\blambda,\tilde{\blambda}}^{(\varrho)}$ with $L$ for $\Delta P_{\rm ph}=\frac{4\pi}{L}, \pi/10, \pi/2$. On the basis of analytic expansions in $1/c$ of matrix elements of $\varrho(x)$, Ref.~\cite{essler2024statistics} conjectured the following form for the form factors \eqref{eq:formfactorrho} associated with $n_{\rm ph}=1$
\begin{equation}
\label{eq:formfacconjecture}
  |F_{\blambda,\tilde{\blambda}}^{(\varrho)}|^2\bigg|_{n_{\rm ph}=1} = \frac{\beta}{L^2}e^{- \alpha \, L} \qquad \rightarrow  \qquad  M_{\blambda,\tilde{\blambda}}^{(\varrho)}\bigg|_{n_{\rm ph}=1} =\alpha \, L + 2 \ln L - \ln \beta \ ,
\end{equation}
where $\alpha$ and $\beta$ are positive $\mathcal{O}(L^0)$ scalars that depend on details of the distribution of rapidities in $\ket \blambda$ and on the particle-hole excitation that leads to $\ket{\tilde{\blambda}}$. To verify this functional form we plot $M_{\blambda,\tilde{\blambda}}^{(\varrho)}-2 \ln L$ as a function of $L$ and find the data to be accurately described by a 2-parameter linear fit $a \, L + b$. For finite momentum transfers $\Delta P_{\rm ph} = \pi/10, \pi/2$ we obtain finite but extremely small (order $10^{-4}$--$10^{-3}$) values of $a$, which estimate the scalar $\alpha$ in \cref{eq:formfacconjecture}. This shows that for system sizes $L$ on the order of a few hundred, \emph{the exponential suppression of these form factors of $\varrho(x)$ is negligible}. In the inset of \cref{fig:scaling}(a) we show an analogous plot for the momentum transfer $\Delta P_{\rm ph}=4 \pi/L$, which vanishes in the thermodynamic limit. Surprisingly, in this case the data are compatible with $\alpha = 0$, i.e.~absence of exponential suppression with $L$. This suggests that in integrable models it is possible to have matrix elements of local operators that are merely polynomially suppressed in $L$ (as in free theories) also in the presence of interactions ($0<c<\infty$). In \cref{fig:scaling}(b) we present the scaling of $M_{\blambda,\bmu}$ with $L$ for the same momentum transfers of \cref{fig:scaling}(a), and for the reference state $\ket{\bmu_{\rm ref}}$ ($\Delta P=0$). The data are well described by 1-parameter linear fits $a \, L$, with $a$ of the order of $10^{-1}$--$10^0$, i.e.~significantly higher than in the case of $M_{\blambda,\tilde{\blambda}}^{(\varrho)}$. This is consistent with the $L$-dependence found in \cref{fig:histograms_T_500_c_7,fig:histograms_T_500_c_70,fig:histograms_T_5_c_7}, and proves that the exponential decay of matrix elements is significantly faster for the Bose field than for the density. Furthermore, it is compatible with the claim that, unlike for the density $\varrho(x)$, \emph{all} matrix elements of the Bose field are at least exponentially suppressed in $L$.

Overall, the numerical results presented yield insights on why it is possible to reconstruct finite-$T$ correlation functions of $\varrho(x)$ by the ABACUS algorithm \cite{caux2009correlation,de_klerk2023improved} up to fairly large values of $N$, of the order of 50--100 (depending on the specific temperature chosen) \cite{panfil2014finite}. By construction, ABACUS can only take into account eigenstates with a few particle hole-excitations on top of $\ket \blambda$, i.e.~$n_{\rm ph}\le n_{\rm max}$ with $n_{\rm max}$ a nonlarge integer threshold. The total number of such eigenstates (given a physical momentum cutoff for their rapidities) scales polynomially with $L$, hence they become eventually irrelevant for the reconstruction of correlation functions at large-enough $L$, due to the exponential decay of the density form factors shown in \cref{fig:scaling}(a) (note that also the form factors that decay only polynomially with $L$, e.g.~those associated with a $\Delta P_{\rm ph}\propto 1/L$, become irrelevant at large $L$ because their total number does not grow with $L$).  However, given the smallness of the coefficient $\alpha$ in their exponential decay $\exp(-\alpha \, N)$, these finite-$n_{\rm ph}$ states remain relevant for the reconstruction of correlation functions at $N \approx 50$--$100$, where  $\alpha \, N \ll 1$ and the exponential suppression is negligible. This is ultimately the reason why ABACUS manages to saturate sum rules for the density $\varrho(x)$ by employing an explicit sum over these type of states. However, similar strategies applied to $\phi(x)$ correlators are limited to significantly smaller temperatures or system sizes \cite{cheng2025one_body}, and further progress requires the use of sampling schemes like the one we presented in the main text. Indeed, already  at $\tau>1$ and $N \lesssim 100$, the total number of eigenstates constructed by applying a small number of particle-hole excitations on top of a reference state is very far from counterbalancing the smallness of their matrix elements, \emph{cf}. \cref{fig:scaling}(b).  

We conclude by remarking that while at larger values of $N$ than those probed in \cite{panfil2014finite}, or at larger temperatures, the use of sampling schemes becomes necessary also for $\varrho(x)$, one cannot directly employ the MCMC strategy described above for the Bose field. Indeed, we showed that a simple proposal update that implements a single particle-hole excitation, coupled with the Metropolis acceptance step, is very successful in exploring the manifold of relevant states. However, due to the lumpiness that characterizes the matrix elements of $\varrho(x)$ (see \cref{fig:histograms_T_500_c_70,fig:histograms_T_500_c_7,fig:histograms_T_5_c_7}), these type of Markov updates are associated with extremely small acceptance rates, which lead to very limited explorations of the landscape of relevant eigenstates. It is an interesting problem for future work to construct an efficient strategy for sampling relevant matrix elements of $\varrho(x)$.


\end{document}